\documentclass[useAMS,usenatbib]{mn2e}
\usepackage{epsfig}
\usepackage{subfigure}
\usepackage{amsmath}
\usepackage{multicol}
\usepackage{threeparttable}

\usepackage{verbatim} 

\title[$^{12}$CO~\textit{J=3$\rightarrow$2} Observations of NGC~2976 and NGC~3351]{The James Clerk Maxwell Telescope Nearby Galaxies Legacy Survey IX: $^{12}$CO~\textit{J=3$\rightarrow$2} Observations of NGC~2976 and NGC~3351}
\author[B. K. Tan et. al.]{Boon-Kok Tan$^{1,2}$\thanks{E-mail:tanbk@astro.ox.ac.uk}, J. Leech$^{1}$, D. Rigopoulou$^{1,3}$, B. E. Warren$^{4,5}$, C. D. Wilson$^{4}$, \newauthor D. Attewell$^{4}$, M. Azimlu$^{6}$, G. J. Bendo$^{7}$, H. M. Butner$^{8}$, E. Brinks$^{9}$, P. Chanial$^{10}$, \newauthor D. L. Clements$^{10}$, V. Heesen$^{9, 11}$, F. Israel$^{12}$, J. H. Knapen$^{13,14}$, H. E. Matthews$^{15}$, \newauthor A. M. J. Mortier$^{10}$, S. M{\"u}hle$^{16}$, J. R. S{\'a}nchez-Gallego$^{13,17}$, R. P. J. Tilanus$^{18,19}$, \newauthor A. Usero$^{8,20}$, P. van der Werf$^{12}$ and M. Zhu$^{21}$\\
$^{1}$Department of Astrophysics, University of Oxford, Keble Road, Oxford, OX1 3RH, UK.\\
$^{2}$Institute for Research and Innovation, Wawasan Open University, 54 Jalan Sultan Ahmad Shah, 10050 Penang, Malaysia.\\
$^{3}$Rutherford Appleton Laboratory, Chilton, Didcot OX11 0QX, UK.\\
$^{4}$Department of Physics \& Astronomy, McMaster University, Hamilton, Ontario L8S 4M1, Canada.\\
$^{5}$Intl. Centre for Radio Astronomy Research, M468, Univ. of Western Australia, 35 Stirling Hwy, Crawley, WA 6009, Australia.\\
$^{6}$Harvard-Smithsonian Center for Astrophysics, 60 Garden St., Cambridge, MA 02138, USA.\\
$^{7}$UK ALMA Regional Centre Node, Jodrell Bank Centre for Astrophysics, Univ. of Manchester, Oxford Rd., Manchester M13 9PL, UK.\\
$^{8}$Department of Physics and Astronomy, James Madison University, MSC 4502-901 Carrier Drive, Harrisonburg, VA 22807, USA.\\
$^{9}$Centre for Astrophysics Research, University of Hertfordshire, College Lane, Hatfield AL10 9AB, UK.\\
$^{10}$Astrophysics Group, Imperial College London, Blackett Laboratory, Prince Consort Road, London SW7 2AZ, UK.\\
$^{11}$School of Physics and Astronomy, University of Southampton, Southampton, SO17 1BJ, UK.\\
$^{12}$Sterrewacht Leiden, Leiden University, P.O. Box 9513, 2300 RA Leiden, The Netherlands.\\
$^{13}$Instituto de Astrof{\'{\i}}sica de Canarias, 38200 La Laguna, Spain.\\
$^{14}$Departamento de Astrof{\'{\i}}sica, Universidad de La Laguna, E-38200, La Laguna, Tenerife, Spain.\\
$^{15}$NRC, Herzberg Institute of Astrophysics, DRAO, P. O. Box 248, White Lake Road, Penticton, British Columbia V2A 69J, Canada.\\
$^{16}$Argelander-Institut f{\H u}r Astronomie, Universit{\H a}t Bonn, Auf dem H{\H u}gel 71,53121 Bonn, Germany.\\
$^{17}$Department of Physics \& Astronomy, University of Kentucky, Lexington, KY 40506-0055, USA.\\
$^{18}$Joint Astronomy Centre, 660 N. A'ohoku Pl., Hilo, HI 96720, USA.\\
$^{19}$Netherlands Organisation for Scientific Research, The Hague, The Netherlands.\\
$^{20}$Observatorio Astron{\'o}mico Nacional, C. Alfonso XII 3, 28014 Madrid, Spain.\\
$^{21}$National Astronomical Observatories, Chinese Academy of Science, 20A Datun Road, Chaoyang District, Beijing, China.\\
}

\begin{document}

\date{Accepted YYYY MMMM DD. Received YYYY MMMM DD; in original form YYYY MMMM DD}

\pagerange{\pageref{firstpage}--\pageref{lastpage}} \pubyear{YYYY}

\maketitle

\label{firstpage}

\begin{abstract}
\label{Abstract}

We present $^{12}$CO~\textit{J=3$\rightarrow$2} maps of NGC~2976 and NGC~3351 obtained with the James Clerk Maxwell Telescope (JCMT), both early targets of the JCMT Nearby Galaxy Legacy Survey (NGLS). We combine the present observations with $^{12}$CO~\textit{J=1$\rightarrow$0} data and find that the computed $^{12}$CO~\textit{J=3$\rightarrow$2} to $^{12}$CO~\textit{J=1$\rightarrow$0} line ratio ($R_{31}$) agrees with values measured in other NGLS field galaxies. We compute the M$_{\rm{H_2}}$ value and find that it is robust against the value of $R_{31}$ used. Using \textsc{Hi} data from the The \textsc{Hi} Nearby Galaxy Survey (THINGS) survey, we find a tight correlation between surface density of H$_2$ and star formation rate density for NGC~3351 when $^{12}$CO~\textit{J=3$\rightarrow$2} data are used. Finally, we compare the $^{12}$CO~\textit{J=3$\rightarrow$2} intensity with the PAH~8~$\mu$m surface brightness and find a good correlation in the high surface brightness regions. We extend this study to include all 25 \textit{Spitzer} Infrared Nearby Galaxies Survey (SINGS) galaxies within the NGLS sample and find a tight correlation at large spatial scales. We suggest that both PAH~8~$\mu$m and $^{12}$CO~\textit{J=3$\rightarrow$2} are likely to originate in regions of active star formation.
\end{abstract}

\begin{keywords}
ISM: molecules, galaxies: individual: NGC~2976, galaxies: individual: NGC~3351, galaxies: ISM, submillimetre, infrared: galaxies
\end{keywords}

\section{Introduction}
\label{Introduction}

Observations of molecular gas are essential for understanding the role of star formation in the evolution of galaxies. Because direct detection of molecular hydrogen (H$_{2}$) is difficult, carbon monoxide (CO) is used as its proxy. Many extra-galactic surveys of low-\textit{J} rotational transitions of CO have been conducted so far \citep[e.g.,][]{Sage:1993, Braine:1993, Young:1995, Elfhag:1996, Meier:2001, Albrecht:2004, Israel:2005}, the majority being single pointing observations that do not provide any information on the spatial distribution of the emission. High angular resolution interferometric surveys have also been conducted \citep[e.g.,][]{Sakamoto:1999, Regan:2001, Helfer:2003}, but most of them have targeted the central regions of the galaxies due to the limited field size. High-sensitivity multi-pixel array receivers (on single-dish telescopes) are now providing much faster speeds for mapping interesting structures in the ISM across entire galactic disks. Two surveys using this type of focal plane array receivers were recently carried out, one by \citet{Leroy:2009} using Heterodyne Receiver Array (HERA) on the Institut de Radio Astronomie Millim{\'e}trique (IRAM) 30~m telescope to map $^{12}$CO~\textit{J=2$\rightarrow$1} and another by \citet{Kuno:2007} using Beam Array Receiver System (BEARS) on Nobeyama Radio Observatory (NRO) to map $^{12}$CO~\textit{J=1$\rightarrow$0}, in nearby galaxies.

The NGLS survey \citep{Wilson:2009, Wilson:2012} uses the JCMT to map $^{12}$CO~\textit{J=3$\rightarrow$2} emission from nearby galaxies. The $^{12}$CO~\textit{J=3$\rightarrow$2} transition traces the warmer and denser regions of the molecular gas that are more directly related to star forming regions \citep[][and references therein]{Wilson:2009}. The entire JCMT NGLS sample consists of 155 nearby galaxies, each with spectral line observations at 345~GHz, made with the Heterodyne Array Receivers Program for B-band (HARP-B) receiver \citep{Smith:2003}. The details of the NGLS survey can be found in \citet{Wilson:2012}.

NGC~2976 and NGC~3351 were observed during the early stages of the NGLS survey. NGC~2976 is a dwarf galaxy on the outskirts of the M81 group, in weak tidal interaction with the group \citep{Appleton:1981, Yun:1999}. Although the galaxy contains primarily an older stellar population, there are indications \citep{Williams:2010} that there are sites of intense star formation at the two bright regions near the north and south ends of the galaxy's major axis \citep{Dale:2009}. NGC~3351 is a starburst galaxy with ring structures \citep{Colina:1997, Elmegreen:1997, Knapen:2002, Swartz:2006, Hagele:2007, Comeron:2013} that displays a very young starburst population. It is thus interesting to understand the effect of such different environments on the physical conditions of the molecular component of the ISM within these two galaxies. Using the improved spatial resolution of the NGLS $^{12}$CO~\textit{J=3$\rightarrow$2} survey, we can resolve the sites of \emph{warmer}, \emph{denser} molecular hydrogen, and investigate the correlation of this observable with other components of the ISM observed in different wavebands. 

Another important motivation for the NGLS is to probe the interplay between dust and gas in nearby galaxies. Polycyclic aromatic hydrocarbons (PAHs) have often been used as an indicator of star formation activity in external galaxies \citep{Zhu:2008, Kennicutt:2009}. An alternative theory that PAH emission is not related to star formation but instead is associated with continuum emission from cold dust has also been proposed by \citet{Haas:2002}. However, results from the \textit{Spitzer} Infrared Nearby Galaxies Survey (SINGS), a comprehensive infrared imaging and spectroscopic survey of 75 nearby galaxies using the \textit{Spitzer} Space Telescope (hereafter \textit{Spitzer}), indicated that the relation of PAHs and star formation extends to spatial scales beyond 1~kpc, although the relation breaks down on smaller spatial scales \citep{Calzetti:2005, Bendo:2006, Calzetti:2007, Prescott:2007, Bendo:2008, Kennicutt:2009, Zhu:2008, Calzetti:2012}. 

In this paper, we report the $^{12}$CO~\textit{J=3$\rightarrow$2} measurements of NGC~2976 and NGC~3351. We describe the basic properties of the two galaxies in Section~\ref{Target Galaxies}. The details of the observations and data reduction are presented in Section~\ref{Observation and Data Reduction}. We present our results and discussion in Section~\ref{Results and Discussion} where we calculate $^{12}$CO~\textit{J=3$\rightarrow$2} to $^{12}$CO~\textit{J=1$\rightarrow$0} line ratios, derive the molecular gas masses and study the correlation between the $^{12}$CO~\textit{J=3$\rightarrow$2} line emission and emission from PAHs. Finally, we summarise our conclusions in Section~\ref{Summary and Conclusions}.

\section{Target Galaxies}
\label{Target Galaxies}

NGC~2976 is an SAB(s:)d peculiar dwarf disk galaxy \citep{Buta:2013} in the M81 group. Single-beam $^{12}$CO~\textit{J=3$\rightarrow$2} and $^{12}$CO~\textit{J=2$\rightarrow$1} observations have been presented by \citet{Israel:2005}, while $^{12}$CO~\textit{J=1$\rightarrow$0} emission has been mapped at high resolution by \citet{Simon:2003} using the Berkeley-Illinois-Maryland Association (BIMA) interferometer. These observations, however, only partially covered the two bright end regions. The present NGLS $^{12}$CO~\textit{J=3$\rightarrow$2} map covers a rectangular area corresponding to $D_{25}/2$, and is therefore wide enough to include both bright complexes.

NGC~3351 is an (R')SB(r,nr)a spiral galaxy \citep{Buta:2013} displaying high-mass star formation in a $15.3''\times11.2''$ circumnuclear ring \citep[hereafter the $15''$ ring;][]{Alloin:1982, Buta:1993, Colina:1997, Elmegreen:1997, Comeron:2010, Comeron:2013}, fuelled by gas accreted through a stellar bar \citep{Swartz:2006}. Most of the NGC~3351 studies have so far focused on this bright central region of the galaxy. However, the optical image \citep{Frei:1996, Abazajian:2009} shown in Fig.~\ref{fig:4in1_3351}~(a) shows a faint ring of $\sim2'$ diameter (hereafter the $2'$ ring) encircling the bar, with signs of spiral arms extending from this ring towards the outermost pseudo-ring feature \citep[major axis diameter $\sim5.5'$, minor axis $\sim3.6'$;][]{Buta:1993}. These features are within the area covered by our $^{12}$CO~\textit{J=3$\rightarrow$2} map of NGC~3351 presented in this paper. 

The general properties of NGC~2976 and NGC~3351 taken from \citet{Vaucouleurs:1991} (unless otherwise stated) are summarised in Table~\ref{tab:galaxyprop}.

\begin{table}
\centering
\caption{General properties of NGC~2976 and NGC~3351 taken from the literature.}
\begin{threeparttable}
\begin{tabular}{lll}
\hline
General Properties 					& NGC~2976 			& NGC~3351		\\[0.5ex]
\hline
Type								& SAB(s:)d pec			& (R')SB(r,nr)a		\\ 
R.A. (J2000)						& 09:47:15.6			& 10:43:58.0		\\ 
Dec (J2000)						& +67:54:50			& +11:42:15		\\ 
Distance (Mpc) \tnote{a}				& 3.56				& 9.33				\\ 
Incl. angle \tnote{b}					& $54^{\circ}$			& $39^{\circ}$		\\
$D_{25}/2$ ($'$) \tnote{c}			& 2.9 $\times$ 1.5		& 3.8 $\times$ 2.2		\\
\hline
\end{tabular}
\begin{tablenotes}
\item Notes:
\item[a] Reference for the distance to NGC~2976 from \citet{Karachentsev:2002} and for NGC~3351 from \citet{Freedman:2001}.
\item[b] Inclination angle from \citet{deBlock:2008}.
\item[c] From \citet{Buta:2007}.
\end{tablenotes}
\end{threeparttable}
\label{tab:galaxyprop}
\end{table}

\subsection{Archival Data} 
\label{Archival Data} 

Both NGC~2976 and NGC~3351 are part of the SINGS survey \citep{Kennicutt:2003} and have rich multi-wavelength ancillary data available. Although a number of $^{12}$CO data sets at various transitions are available, most of these are single-beam data and thus the comparison with our $^{12}$CO~\textit{J=3$\rightarrow$2} maps is challenging. Furthermore, the spatial distribution information in our high-resolution $^{12}$CO~\textit{J=3$\rightarrow$2} map will also be under-utilised if used together with these single-beam data. Hence, in this paper, we shall discard the use of single-beam data from the literature, but focus only on the available $^{12}$CO~\textit{J=1$\rightarrow$0} maps. For NGC~3351, $^{12}$CO~\textit{J=1$\rightarrow$0} maps are available from the single-dish NRO \citep{Kuno:2007}\footnote{http://www.nro.nao.ac.jp/$\sim$nro45mrt/COatlas/} and the BIMA Survey of Nearby Galaxies \citep[SONG;][]{Helfer:2003}. We used $^{12}$CO~\textit{J=1$\rightarrow$0} maps from NRO as their beam size ($15''$) closely matches the $^{12}$CO~\textit{J=3$\rightarrow$2} beam size of HARP-B on the JCMT. Unfortunately NGC~2976 was not included in the NRO survey. Hence, the $^{12}$CO~\textit{J=1$\rightarrow$0} map for NGC~2976 was retrieved from the BIMA SONG survey\footnote{http://nedwww.ipac.caltech.edu/level5/March02/SONG/\\SONG.html}. However, we have not used this map for any further analysis, but have only displayed the $^{12}$CO~\textit{J=1$\rightarrow$0} distribution within the galaxy in Figure~\ref{fig:4in1_2976}, as the BIMA $^{12}$CO~\textit{J=1$\rightarrow$0} map for NGC~2976 has a poor uv-plane sampling.

The \textit{Spitzer} Infrared Array Camera (IRAC) 3.6~$\mu$m and 8~$\mu$m data and Multiband Imaging Photometer for \textit{Spitzer} (MIPS) 24~$\mu$m data used in this study were downloaded from the SINGS survey website\footnote{http://sings.stsci.edu/}. The optical images for both galaxies were retrieved from the seventh data release of the Sloan Digital Sky Survey (SDSS)\footnote{http://www.sdss.org/}. The \textsc{Hi} images were downloaded from the THINGS survey\footnote{http://www.mpia.de/THINGS/Data.html} and the far ultraviolet (FUV) images used to produce the star formation rate (SFR) surface density images, in conjunction with the  24~$\mu$m data, were retrieved from the Galaxy Evolution Explorer ({\em GALEX}) data release\footnote{http://galex.stsci.edu/GR4/}.

\section{Observations \& Data Reduction}
\label{Observation and Data Reduction}

The $^{12}$CO~\textit{J=3$\rightarrow$2} (rest frequency 345.796~GHz) observations for NGC~2976 were carried out over two nights between November 2007 and January 2008, while observations for NGC~3351 took place over two runs in January 2008. The instrument used was HARP-B which has 16 Superconductor-Insulator-Superconductor (SIS) heterodyne mixers arranged in a 4$\times$4 array with $30''$ row and column separation. This corresponds to a $2'$ square footprint on the sky. HARP-B operates over a frequency range of 325~--~375~GHz and the average Full Width Half Maximum (FWHM) beam width is $14.5''$. The receiver operates with the Auto-Correlation Spectrometer and Imaging System \citep[ACSIS;][]{Buckle:2009} as the back-end data processing unit. The observations for both galaxies in this paper were made using a 1~GHz ACSIS bandwidth with a spectral resolution of 0.488~MHz. The main beam efficiencies ($\eta_{\rm MB}$) used to convert the corrected antenna temperature ($T_{\rm A}^*$) to the main beam brightness temperature ($T_{\rm mb}$) were determined from observations of bright planets. All data presented in this paper were calibrated to the $T_{\rm mb}$ scale using $\eta_{\rm MB}=0.6$.

\begin{table}
\caption{Observation dates and set-up used in the observing runs for NGC~2976 and NGC~3351.}
\centering
\begin{minipage}{90mm}
\begin{threeparttable}
\begin{tabular}{lll}
\hline
Parameters	 		& NGC~2976 			& NGC~3351	\\[0.5ex]
\hline
Observing dates		& 25 Nov 2007		& 07 Jan 2008	\\
					& 06 Jan 2008		& 12 Jan 2008	\\
Pos. angle (major axis)		& 143$^\circ$		& 13$^\circ$ 	\\ 
Height of map 		& $294''$			& $348''$		\\ 
Width of map 		& $210''$			& $252''$		\\ 
$T_{\rm sys}$ 			& 341 K				& 409 K	\\
$\Delta T^*_{\rm A}$		& 13.3 mK			& 16.2 mK	\\ 
Pixel size			& $7.2761''$			& $7.2761''$ \\
\hline
\end{tabular}
\end{threeparttable}
\label{tab:obsdetail}
\end{minipage}
\end{table}

Both NGC~2976 and NGC~3351 were raster scanned using a basket-weave technique with half array steps ($58.2''$, the reader is referred to the appendix in \citet{Warren:2010} for a more detailed description of the steps used in the data reduction process). This ensured that all of the area within the target scan region, defined to include all of the optical galactic emission, was fully sampled. These fully sampled maps were made repeatedly until the target root-mean-square (RMS) noise of the combined scans (less than 19~mK in $T_{\rm A}^*$ scale) was achieved within a frequency bin of 20~km~s$^{-1}$ resolution. The observational details for each galaxy are summarised in Table~\ref{tab:obsdetail}.

\subsection{Data Reduction}
\label{Data Reduction}

The spectral data reduction and analysis was done mainly using the \textit{Starlink}\footnote{\textit{Starlink} is maintained by the Joint Astronomy Centre (JAC) (http://www.starlink.ac.uk)} software packages \citep{Jenness:2009}. We used Kernel Application Package (KAPPA)\footnote{http://docs.jach.hawaii.edu/star/sun95.htx/sun95.html} and Sub-Millimetre User Reduction Facility (SMURF) applications within \textit{Starlink} as the main reduction tools. Graphical Astronomy and Image Analysis Tool (GAIA) and Spectrum Analysis Tool (SPLAT) were used for analysis and visualisation purposes. Data collected under the NGLS survey were processed primarily following the steps outlined in \citet{Warren:2010}, with some modifications depending on the characteristics of the individual galaxy and the quality of the observed data \citep{Wilson:2012}.

\subsubsection{$R_{31}$ Line Ratio}
\label{Line Ratio}

To derive the $R_{31}$ map for NGC~3351 using the NRO's $^{12}$CO~\textit{J=1$\rightarrow$0} map, we re-gridded the $^{12}$CO~\textit{J=1$\rightarrow$0} map to match our pixel size ($7.28''$) and calculated the $R_{31}$ ratio map by performing a pixel-by-pixel division of the resulting $^{12}$CO~\textit{J=3$\rightarrow$2} and $^{12}$CO~\textit{J=1$\rightarrow$0} maps. We did not convolve and match the beam of both maps because the beam sizes of NRO (15$''$) and HARP-B (14.5$''$) are very similar. As explained in Section~\ref{Archival Data}, the BIMA $^{12}$CO~\textit{J=1$\rightarrow$0} map for NGC~2976 has poor uv-plane sampling, hence we have not produced the $R_{31}$ map for NGC~2976.

\subsubsection{8~$\mu$m Data \& Radial Profile}
\label{8um Data and Radial Profile}

The point spread function (PSF) of \textit{Spitzer} images is highly non-Gaussian. We thus created convolution kernels, following recipes from \citet{Gordon:2008} and \citet{Bendo:2010}, to match the PSFs to those of HARP-B. These kernels were created, for each waveband, using STinyTim\footnote{http://ssc.spitzer.caltech.edu/archanaly/contributed/browse.html} \citep{Krist:2002}, and convolved with the \textit{Spitzer} images to match the HARP-B beam. We used the KAPPA routine \texttt{convolve} for this task. Further details of the PSF matching for HARP-B and \textit{Spitzer} images can be found in \citet{Bendo:2010}.

The originally reduced IRAC 8~$\mu$m data from the SINGS survey sampled emission from both stars and dust. To produce an 8~$\mu$m image with surface brightness due to dust only (hereinafter dust-only 8~$\mu$m image), we needed to remove the stellar contribution using the 3.6~$\mu$m image \citep{Helou:2004, Smith:2007} following the steps outlined in \citet{Bendo:2010}. First, we determined and subtracted the residual background of both the 3.6~$\mu$m and 8~$\mu$m images by fitting a smoothed gradient of the background brightness outside the galaxy disk. Regions with 3.6~$\mu$m to 8~$\mu$m surface brightness ratio $\geq$5 were masked out as bright foreground stars. The effective aperture corrections were then applied to both images by multiplying the correction factors, 0.944 for the 3.6~$\mu$m image and 0.737 for the 8~$\mu$m image respectively, following the calibration recommendation in \citet{Reach:2005}. Finally, we subtracted the stellar continuum (represented by the final 3.6~$\mu$m image) from the 8~$\mu$m surface brightness images \citep{Helou:2004} using
\begin{equation}
I_{8~\mu \rm{m}}^{\rm{dust-only}} = I_{8~\mu \rm{m}}^{\rm{raw}} - 0.232I_{3.6~\mu \rm{m}}^{\rm{raw}}
\end{equation}
\noindent where $I_{8~\mu \rm{m}}^{\rm{raw}}$ and $I_{3.6~\mu \rm{m}}^{\rm{raw}}$ are the raw 8~$\mu$m and 3.6~$\mu$m intensity map from the IRAC pipeline, and $I_{8~\mu \rm{m}}^{\rm{dust-only}}$ is the final dust-only 8~$\mu$m intensity map. Note that the IRAC 8~$\mu$m image of NGC~3351 was affected by the muxbleed artefact \citep{Laine:2011}, so this area has been masked out.

To create a radial profile, we binned the corresponding maps into a number of elliptical annuli. The ellipticity of the annulus was defined by the ratio of the galaxy's major and minor axis lengths. The width of the annulus was about $14.5''$ for $^{12}$CO~\textit{J=3$\rightarrow$2} data and $5.25''$ for PAH 8~$\mu$m data, defined along the major axis of the galaxy. The radial surface brightness was then the average of the brightness within each ellipse.

\subsubsection{SFR Surface Density}
\label{SFR Surface Density}

For the SFR surface density maps, we combined the {\em GALEX} FUV data with the \textit{Spitzer} 24~$\mu$m maps. The FUV samples the photospheric emission from the O and B stars which relates to the unobscured star formation, whereas the 24~$\mu$m flux traces the emission from dust emission heated by the young stars embedded within. We estimated the SFR surface density using
\begin{equation}
\Sigma_{\rm SFR} = 8.1\times10^{-2}I_{\rm FUV} + 3.2\times10^{-3}I_{24},
\label{eq:SFR}
\end{equation}
from \citet{Leroy:2008}, where $\Sigma_{\rm SFR}$ is the estimated SFR surface density having units of $M_{\odot}$~kpc$^{-2}$~yr$^{-1}$ and both FUV and 24~$\mu$m intensity are in MJy~sr$^{-1}$. We refer the reader to the appendix of \citet{Leroy:2008} for details on the calibration steps.

\section{Results \& Discussion}
\label{Results and Discussion}

\begin{figure*}
\centering
\subfigure[Optical (SDSS, 1.35$''$)]{
\includegraphics*[width=3in, angle=0]{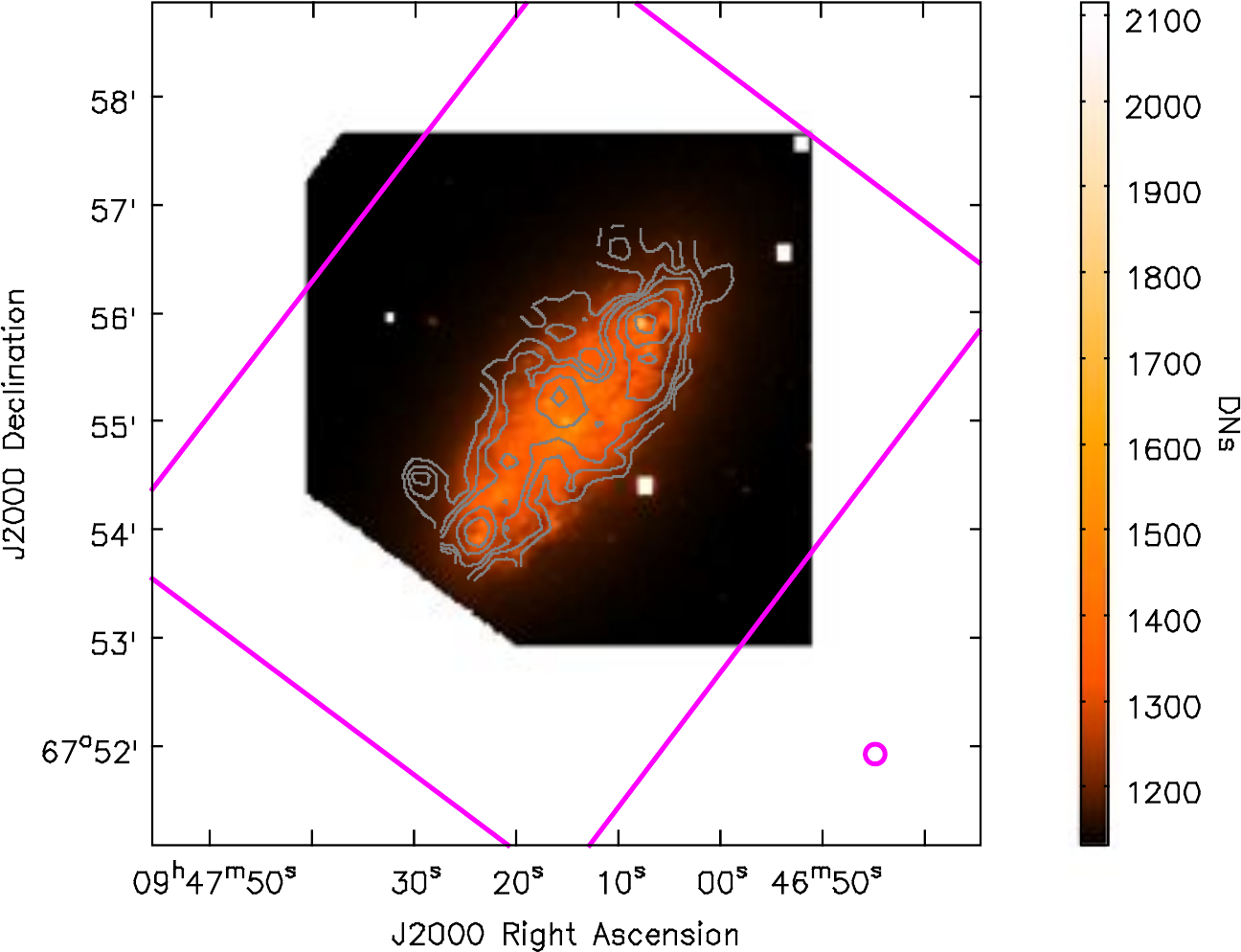}}
\hspace{0.2in}
\subfigure[8~$\mu$m (\textit{Spitzer} IRAC, 0.75$''$)]{
\includegraphics*[width=3in, angle=0]{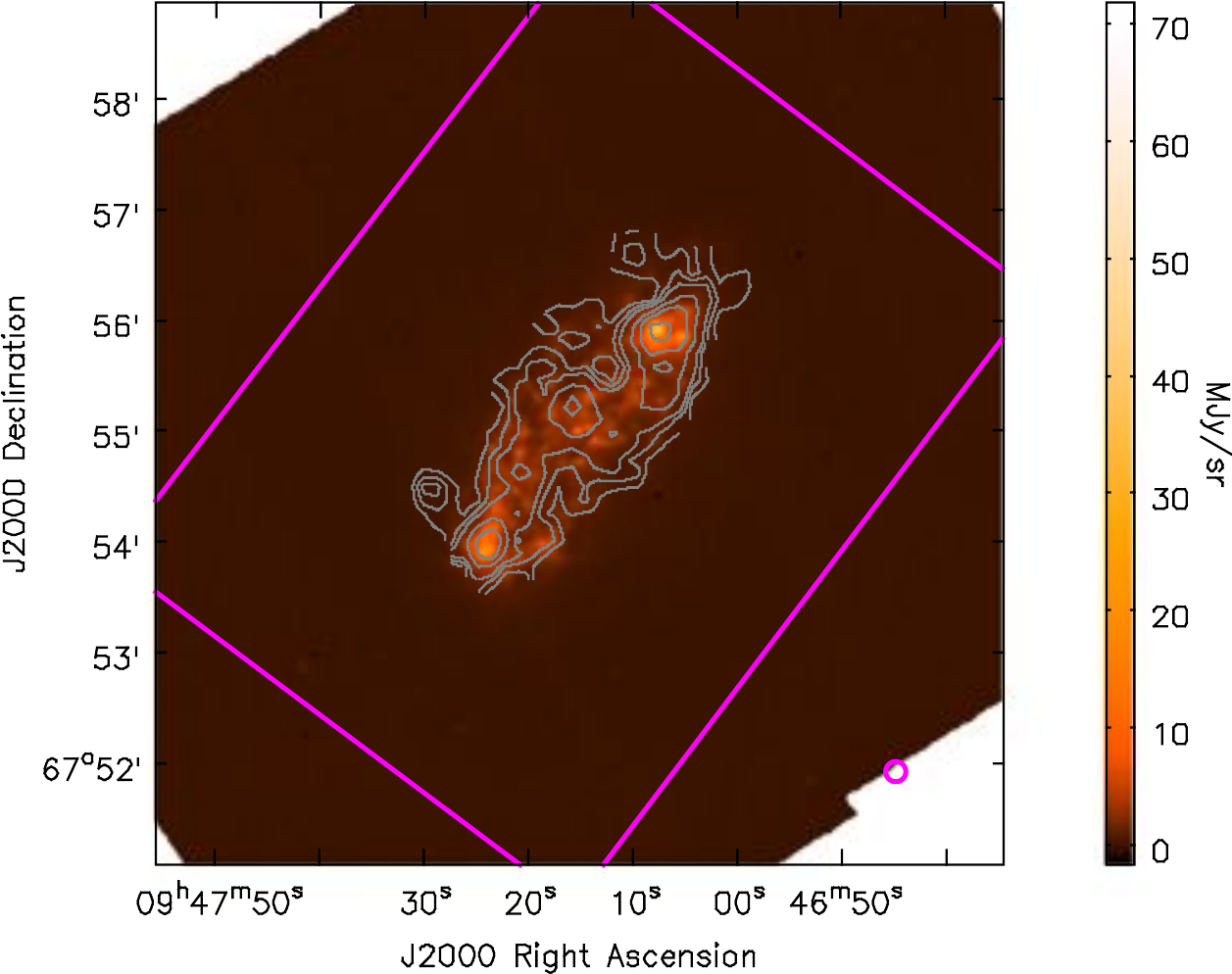}}
\vspace{0.2in}
\subfigure[$^{12}$CO~\textit{J=1$\rightarrow$0} (BIMA, 1$''$)]{
\includegraphics*[width=3in, angle=0]{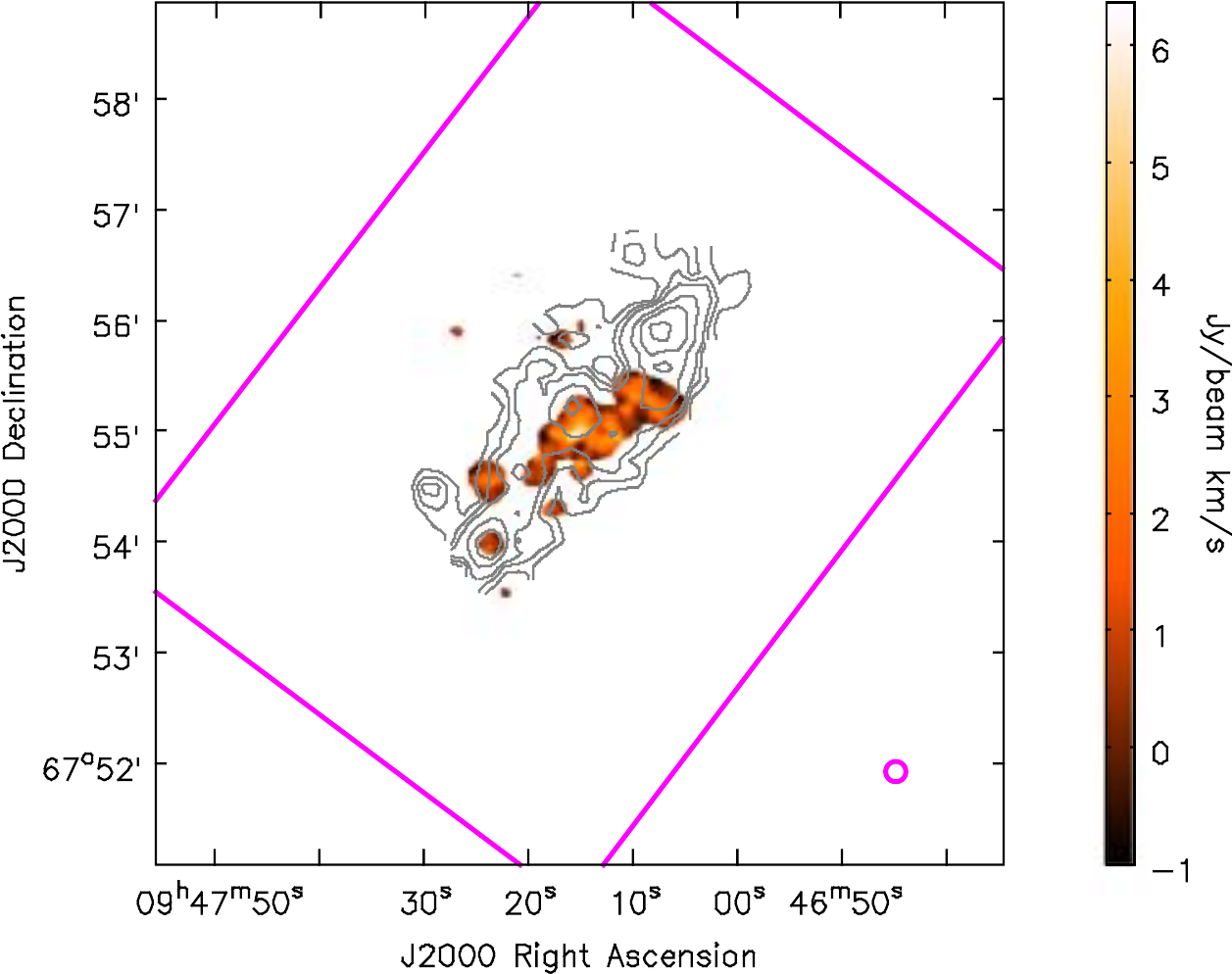}}
\hspace{0.2in}
\subfigure[$^{12}$CO~\textit{J=3$\rightarrow$2} (JCMT NGLS, 7.27$''$)]{
\includegraphics*[width=3in, angle=0]{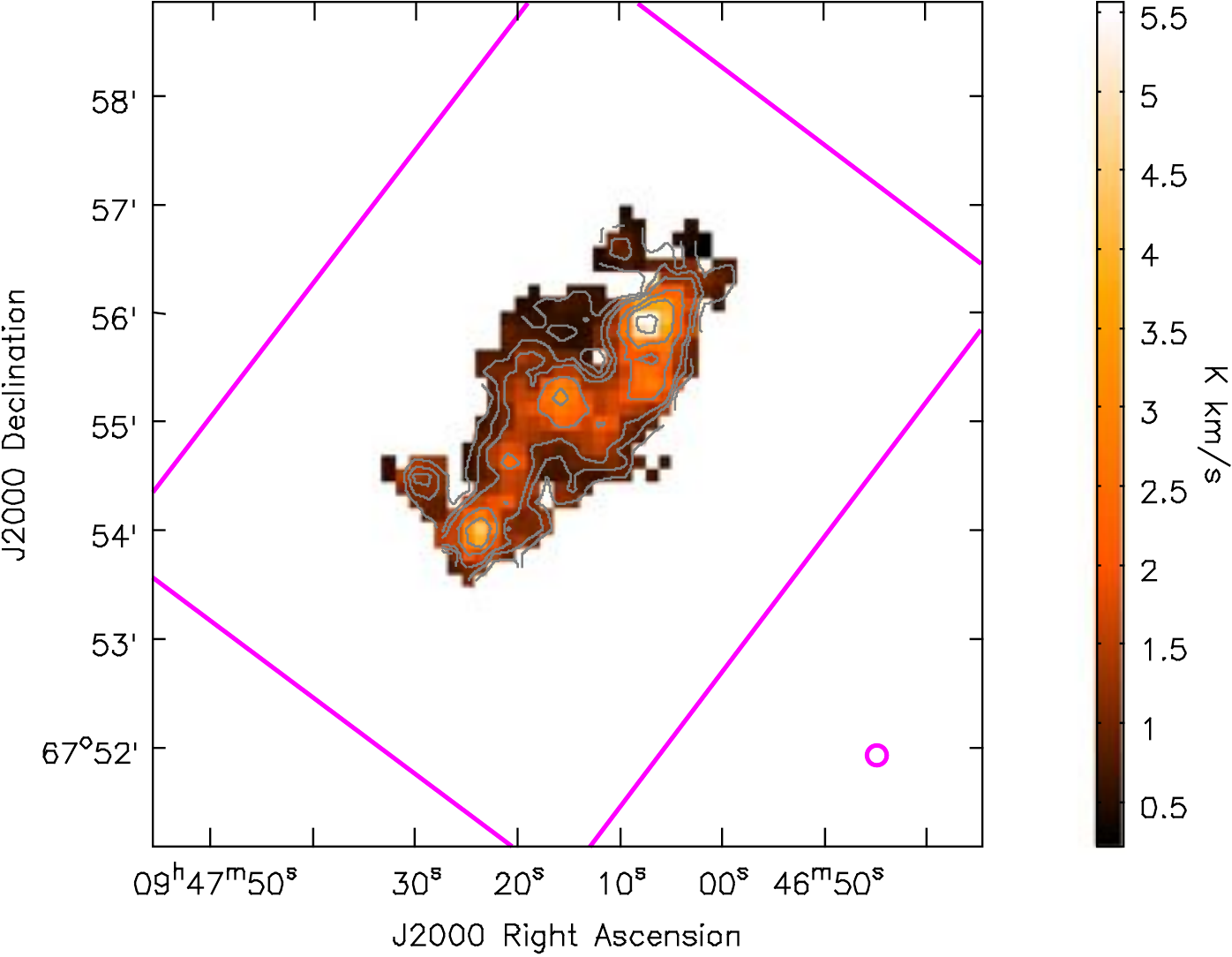}}
\vspace{0.2in}
\subfigure[$\Sigma_{\rm SFR}$ (\textit{Spitzer} MIPS 24~$\mu$m \& \textit{GALEX} FUV, 1.5$''$)]{
\includegraphics*[width=3in, angle=0]{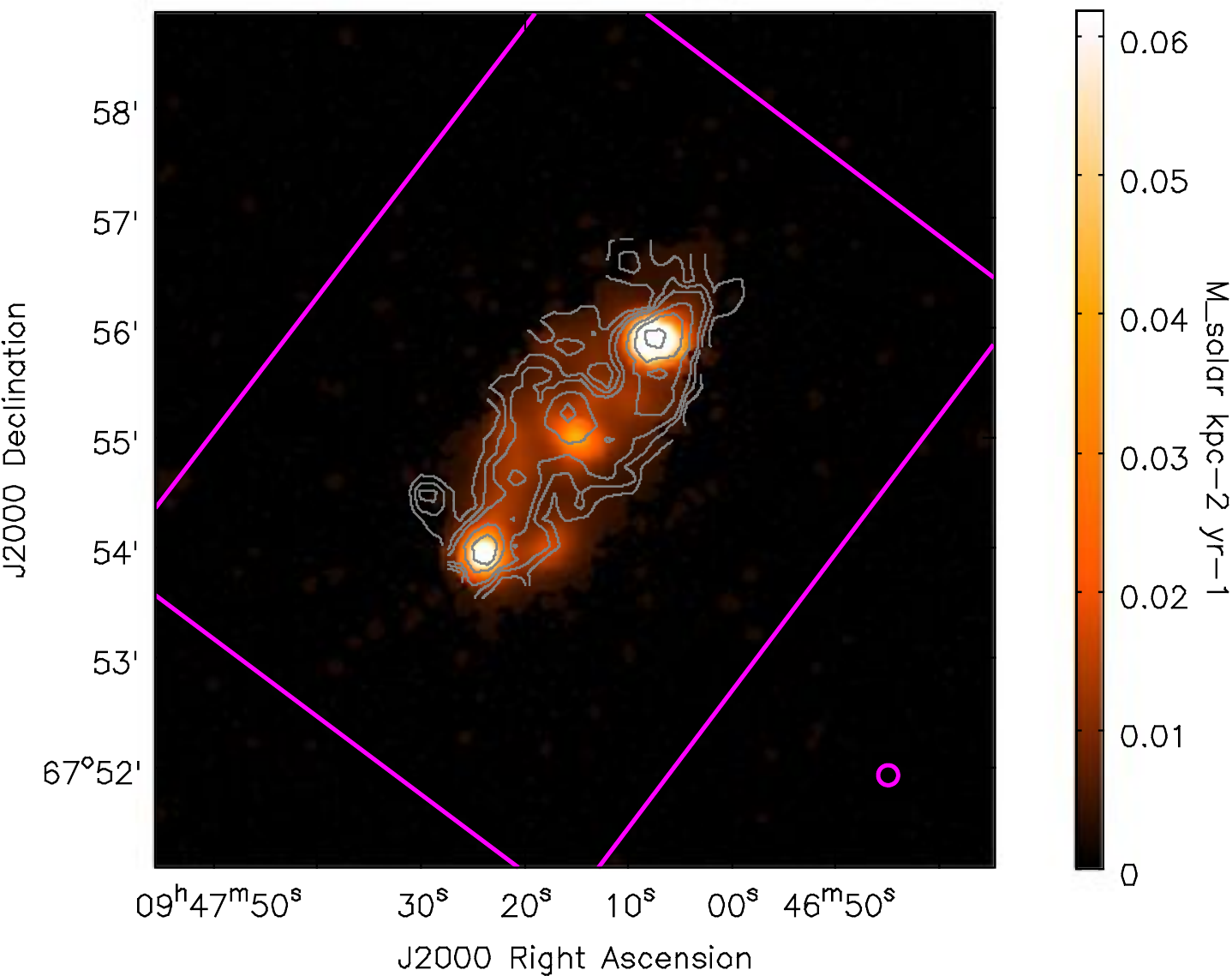}}
\hspace{0.2in}
\subfigure[\textsc{Hi} (THINGS, 1.5$''$)]{
\includegraphics*[width=3in, angle=0]{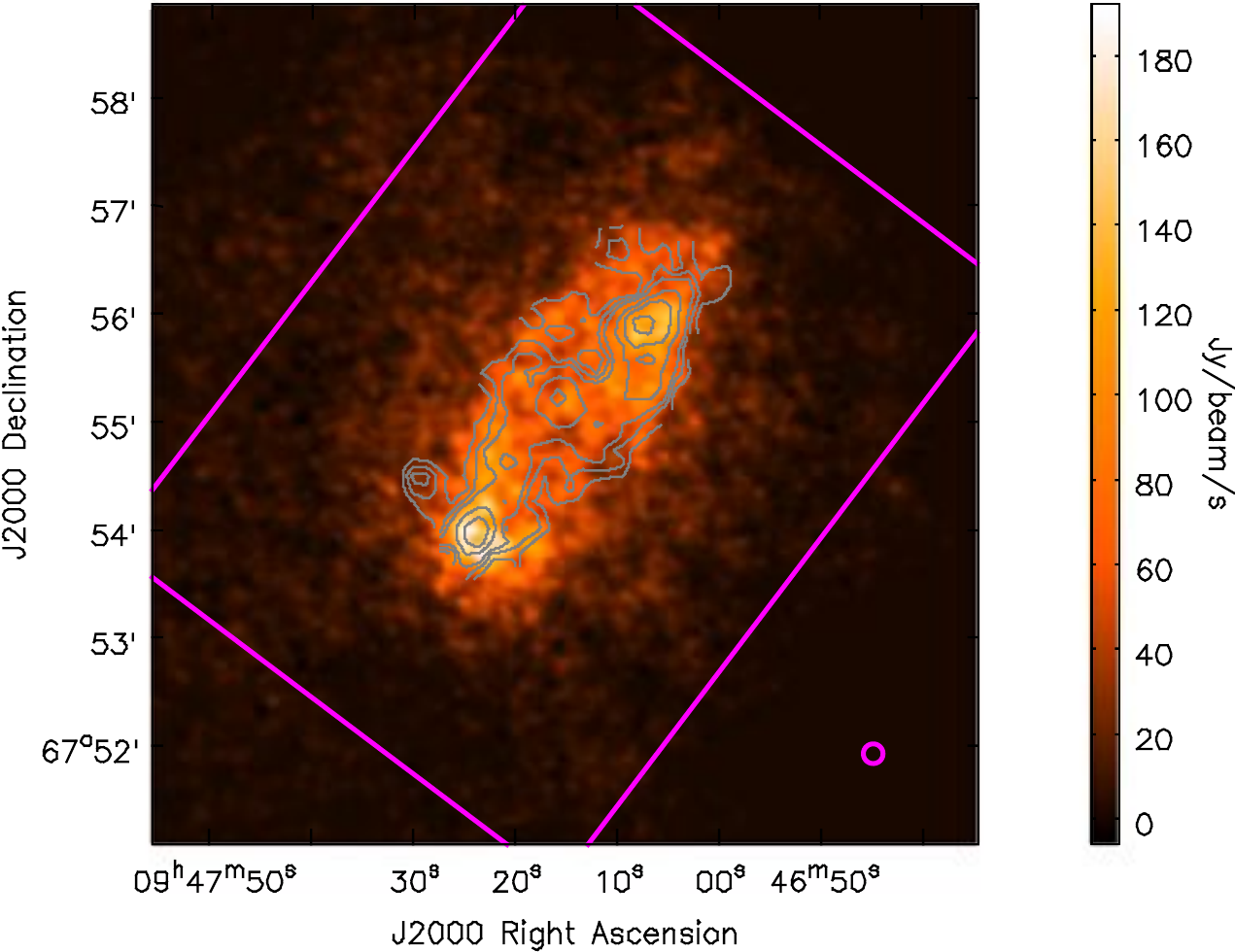}}
\caption{Contours of the $^{12}$CO~\textit{J=3$\rightarrow$2} data of NGC~2976 overlaid on the corresponding ancillary images from the archives. The contour levels are 0.36 (3$\sigma$), 0.72, 1.08, 2.0, 3.0 and 5.0~K~km~s$^{-1}$ (temperature in $T_{\rm mb}$). In all panels, a magenta box is drawn to show the region mapped by HARP-B, and a magenta circle to indicate the 14.5$''$ angular resolution of the $^{12}$CO~\textit{J=3$\rightarrow$2} data. All images are oriented North up and East to the left. The image representing $\Sigma_{\rm SFR}$ has been convolved to the HARP-B beam size. The native resolution of each map is listed along with the title of each individual panels.}
\label{fig:4in1_2976}
\end{figure*}

\begin{figure*}
\centering
\subfigure[Optical (SDSS, 1.35$''$)]{
\includegraphics*[width=3in, angle=0]{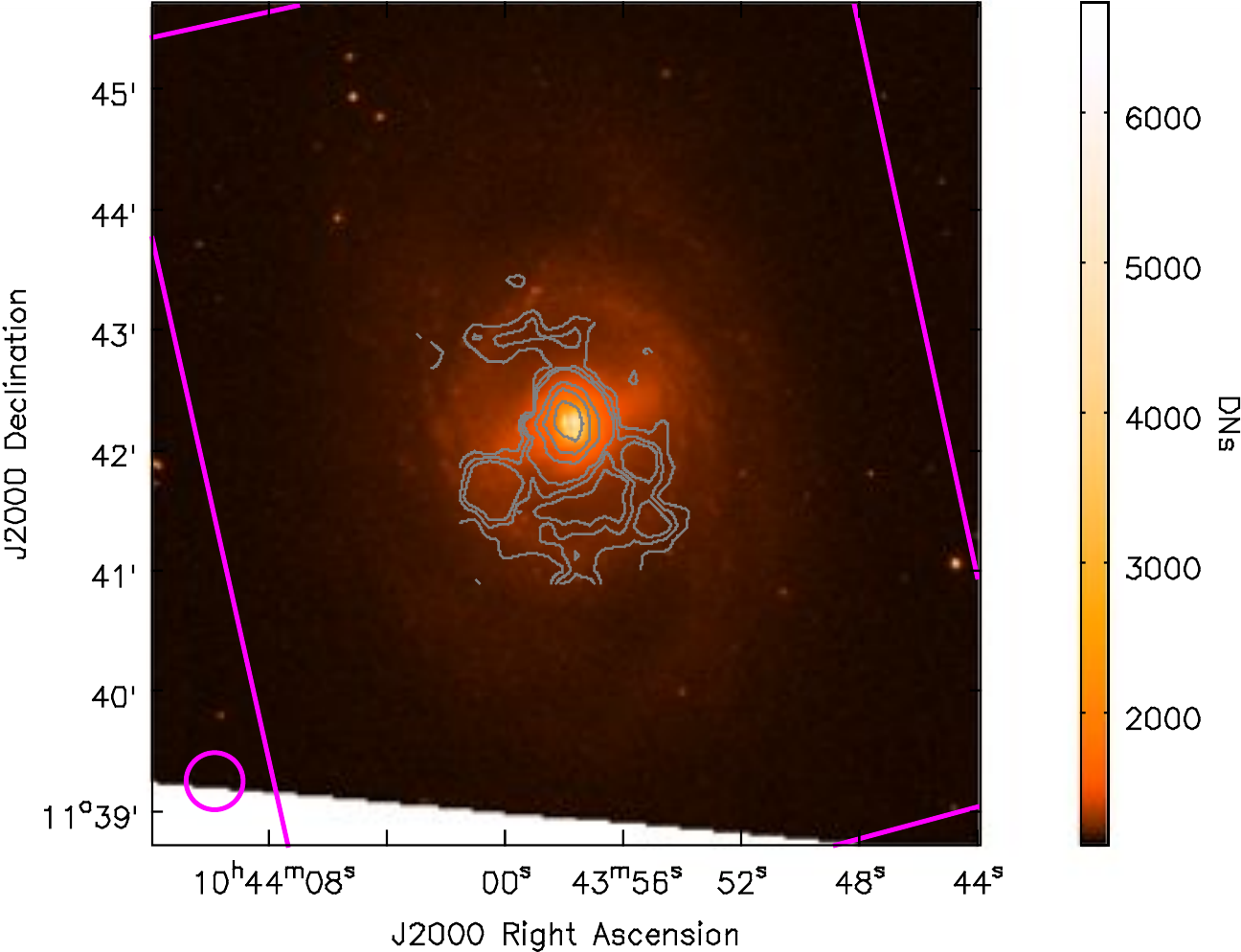}}
\hspace{0.2in}
\subfigure[8~$\mu$m (\textit{Spitzer} IRAC, 0.75$''$)]{
\includegraphics*[width=3in, angle=0]{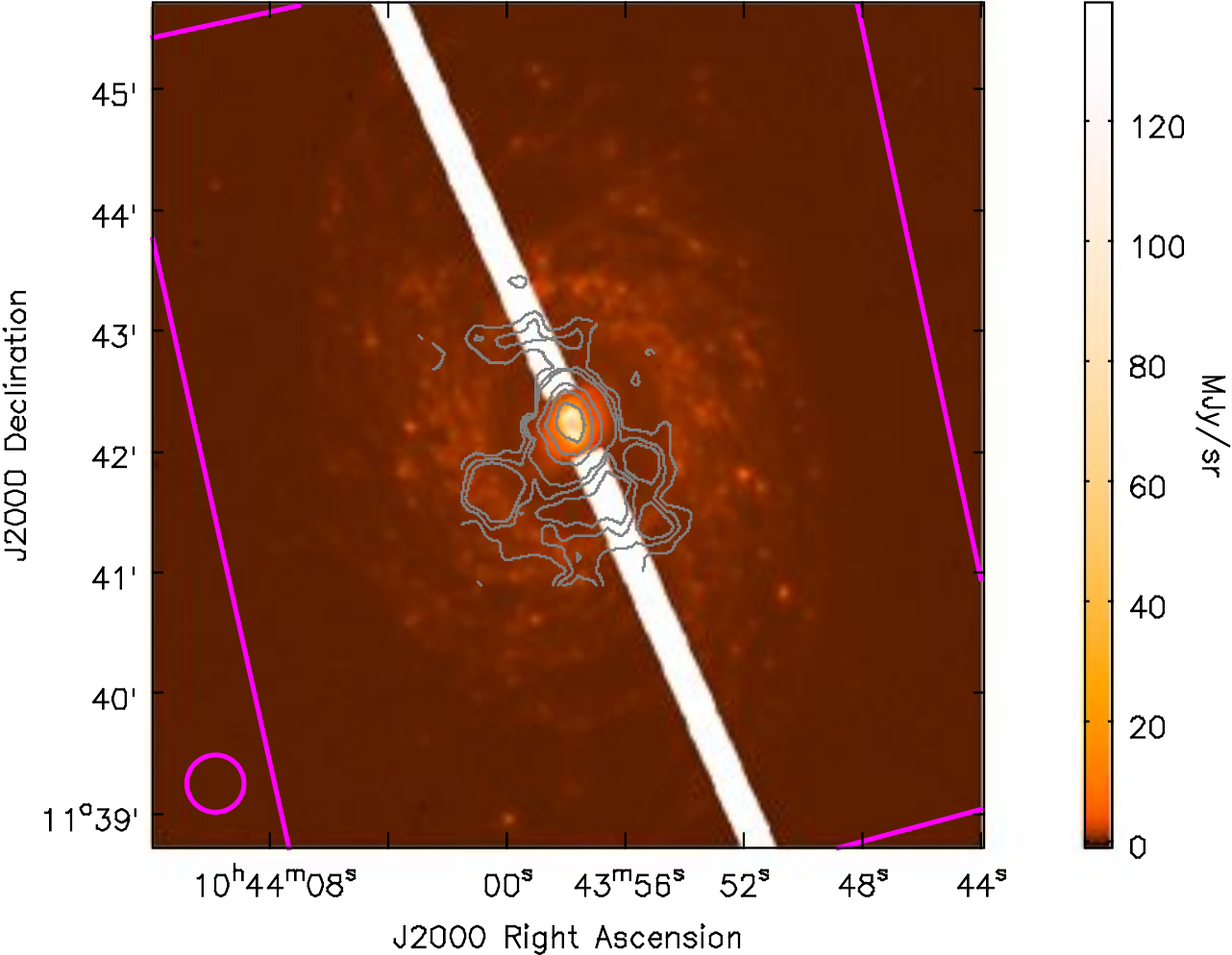}}
\vspace{0.2in}
\subfigure[$^{12}$CO~\textit{J=1$\rightarrow$0} (NRO, 1$''$)]{
\includegraphics*[width=3in, angle=0]{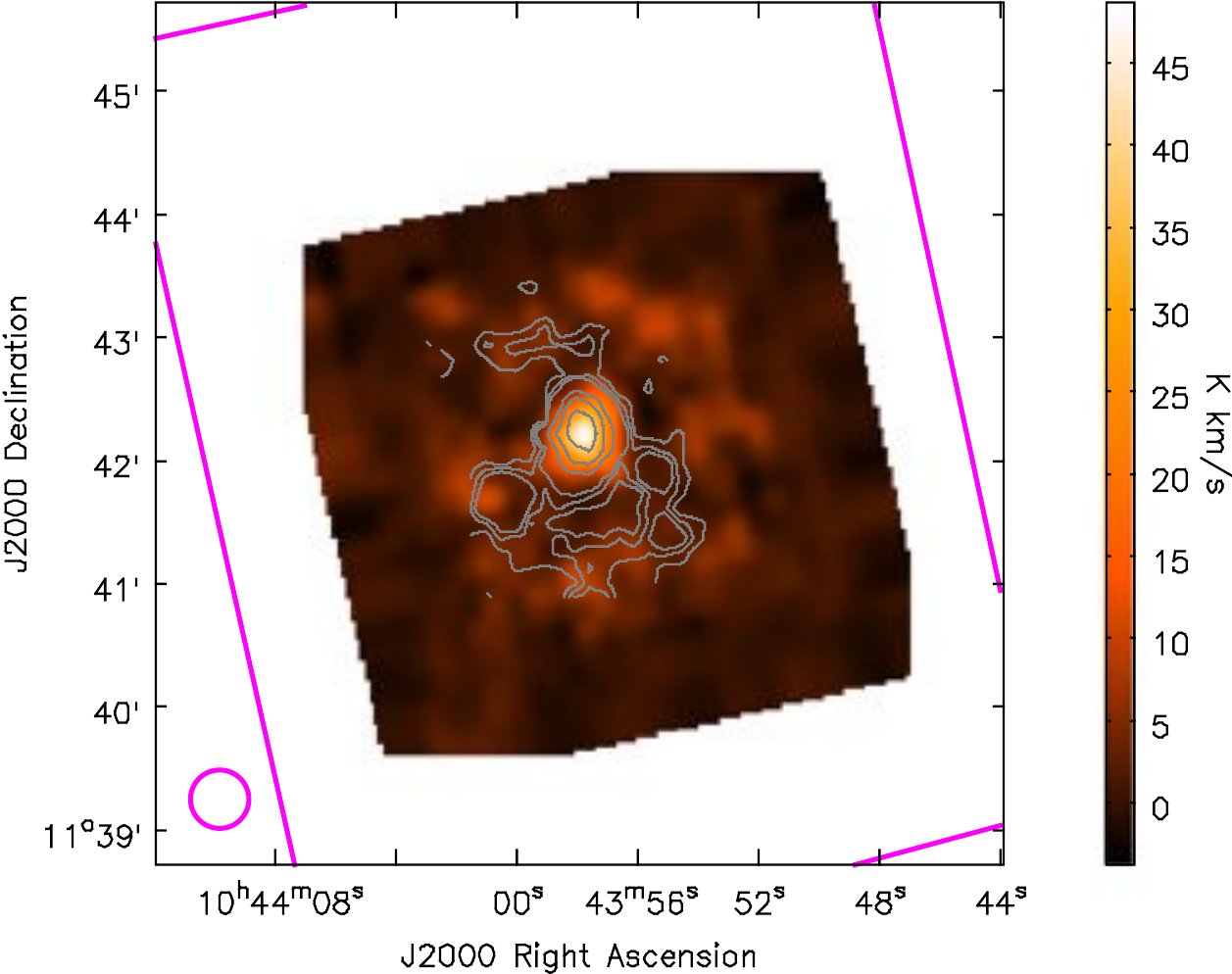}}
\hspace{0.2in}
\subfigure[$^{12}$CO~\textit{J=3$\rightarrow$2} (JCMT NGLS, 7.27$''$)]{
\includegraphics*[width=3in, angle=0]{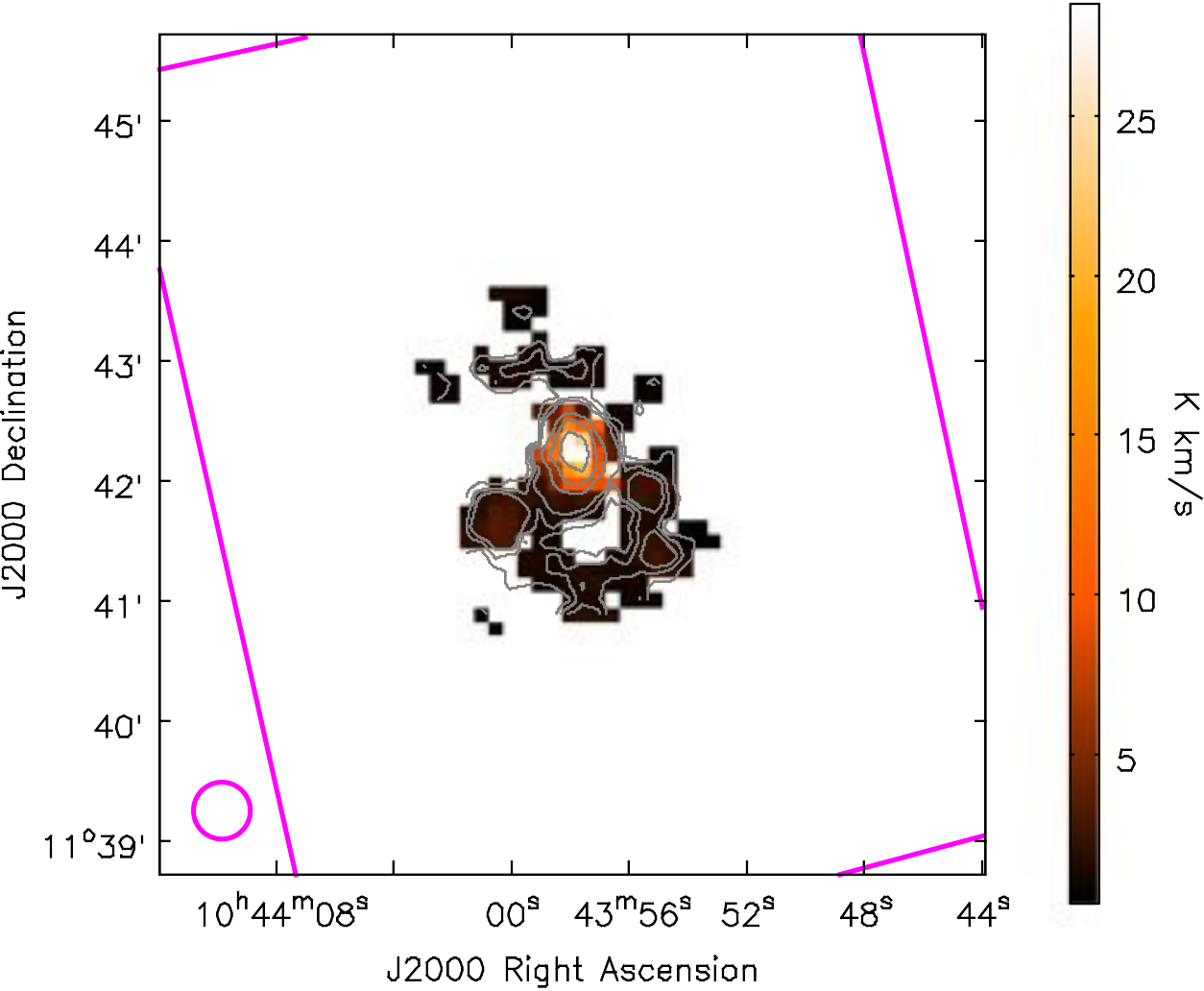}}
\vspace{0.2in}
\subfigure[$\Sigma_{\rm SFR}$ (\textit{Spitzer} MIPS 24~$\mu$m \& \textit{GALEX} FUV, 1.5$''$)]{
\includegraphics*[width=3in, angle=0]{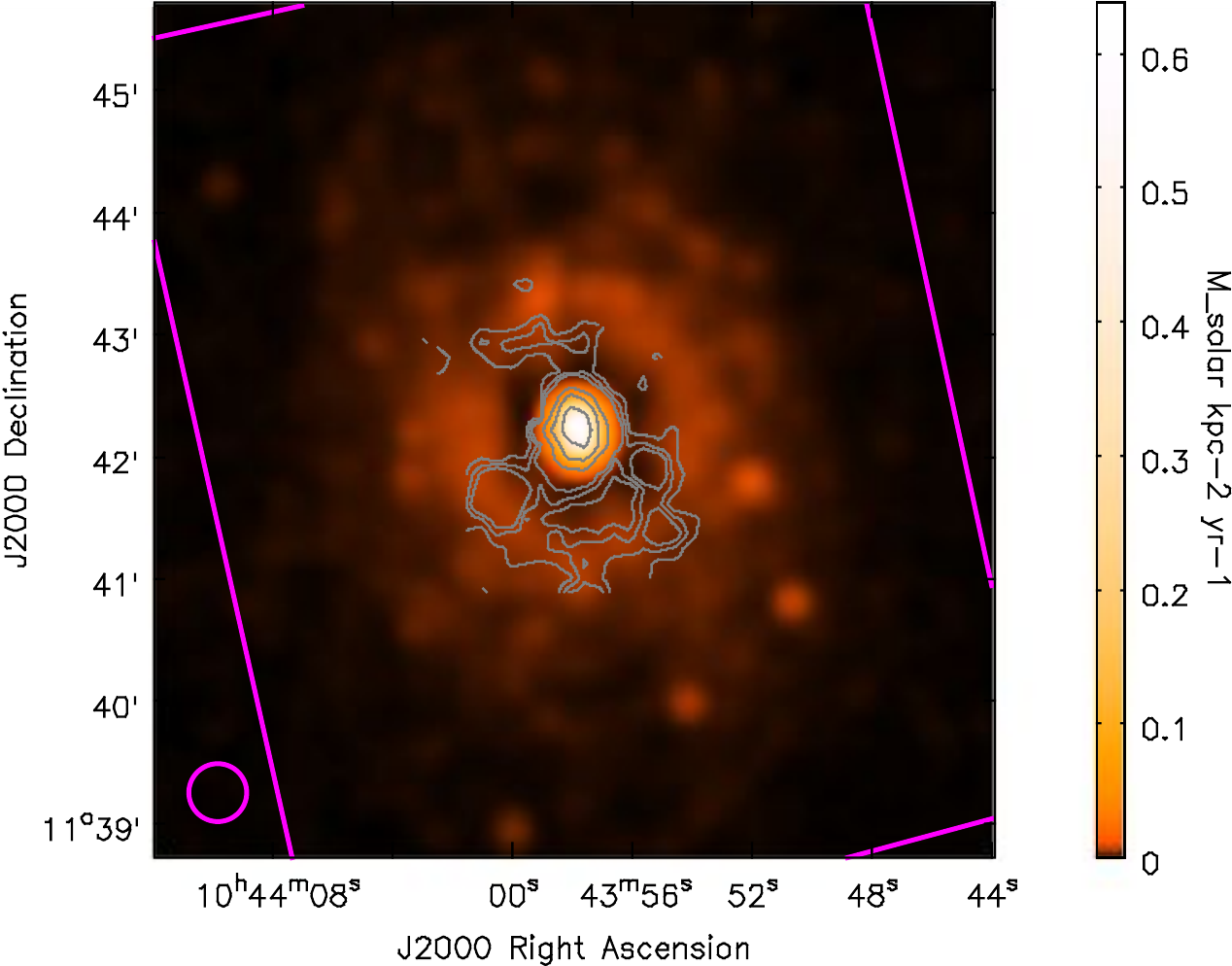}}
\hspace{0.2in}
\subfigure[\textsc{Hi} (THINGS, 1.5$''$)]{
\includegraphics*[width=3in, angle=0]{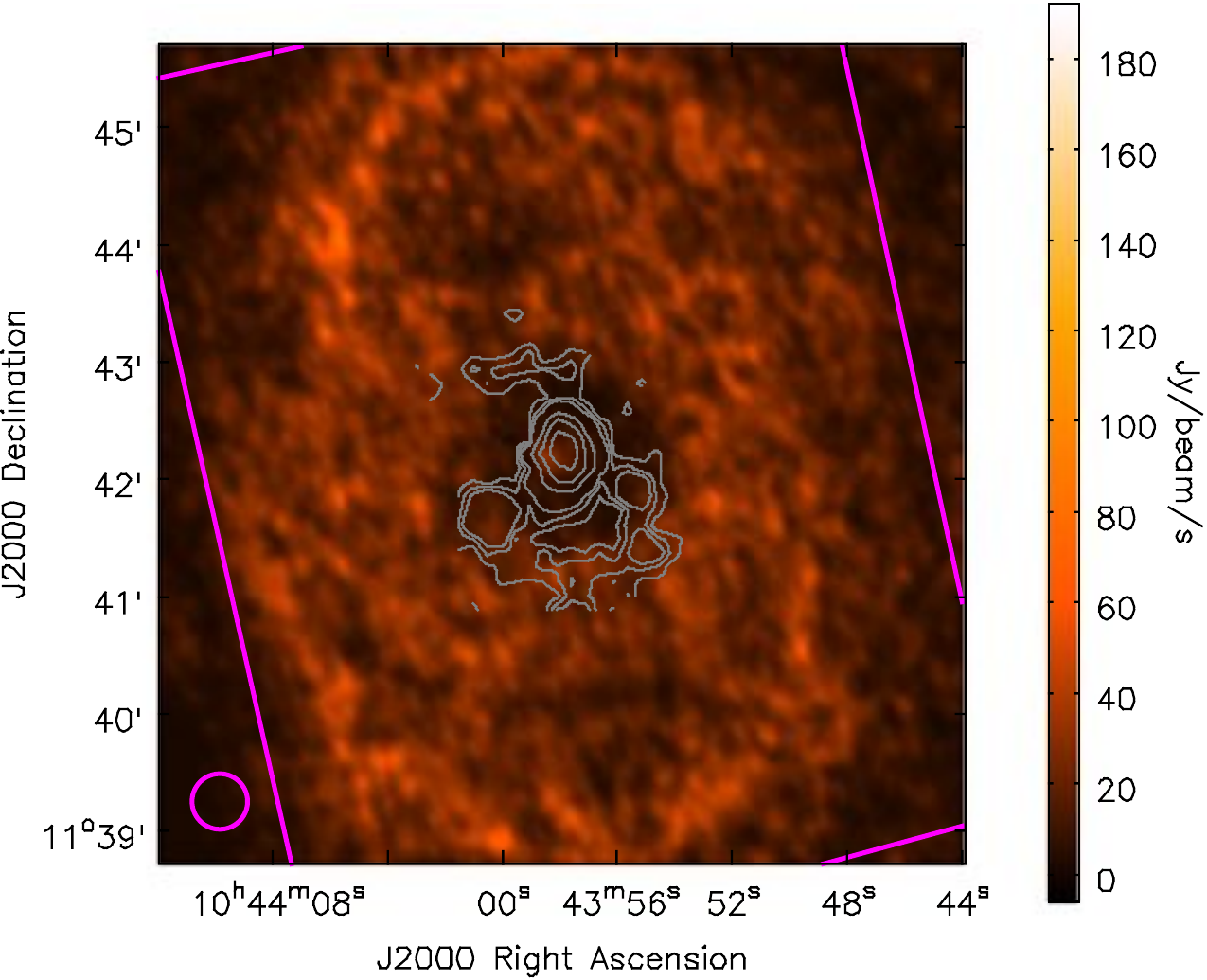}}
\caption{Contours of the $^{12}$CO~\textit{J=3$\rightarrow$2} data of NGC3351 overlaid on the corresponding ancillary images from the archives. The contour levels are 0.43 (3$\sigma$), 0.86, 1.29, 5.0, 10.0, 20.0 and 30.0~K~km~s$^{-1}$ (temperature in $T_{\rm mb}$). In all panels, a magenta box is drawn to show the region mapped by HARP-B, and a magenta circle to indicate the 14.5$''$ angular resolution of the $^{12}$CO~\textit{J=3$\rightarrow$2} data. All images are oriented North up and East to the left. The 8~$\mu$m image is affected by muxbleed \citep{Laine:2011}, hence the affected area is blanked out. The image representing $\Sigma_{\rm SFR}$ has been convolved to the HARP-B beam size. The native resolution of each map is listed along with the title of each individual panels.}
\label{fig:4in1_3351}
\end{figure*}

In Figs.~\ref{fig:4in1_2976} and \ref{fig:4in1_3351}, we show the reduced $^{12}$CO~\textit{J=3$\rightarrow$2} maps of NGC~2976 and NGC~3351, respectively, together with the ancillary maps of $^{12}$CO~\textit{J=1$\rightarrow$0}, IRAC 8~$\mu$m, optical image from SDSS, \textsc{Hi} image from THINGS and SFR surface density map. All maps shown are in their native resolution (except the $\Sigma_{\rm SFR}$ map that was convolved to the HARP-B beam size), overlaid with the contours of our NGLS $^{12}$CO~\textit{J=3$\rightarrow$2} data. 

The NGC~2976 $^{12}$CO~\textit{J=3$\rightarrow$2} map traces an inverse-S-like feature along the major axis. The structure is not seen in the other wavebands discussed here but does exist in the $^{12}$CO~\textit{J=2$\rightarrow$1} image from the HERA CO-Line Extragalactic Survey \citep[HERACLES;][]{Leroy:2009}. This is possibly caused by the lower resolution of the CO maps compared to images at other wavebands. This results in small scale structures being smoothed out in the CO map, hence, making the large structure easier to identify. In both the $^{12}$CO~\textit{J=2$\rightarrow$1} and our $^{12}$CO~\textit{J=3$\rightarrow$2} maps, the emission near the two ends of the major axis is stronger in comparison to the central region. The same strong emission is detected in the IRAC 8~$\mu$m, MIPS 24~$\mu$m and THINGS \textsc{Hi} observations but not in the optical images.

A weak detection near the central region of our $^{12}$CO~\textit{J=3$\rightarrow$2} map is only evident in the SFR surface density image, indicating the existence of hot dust in this region. We note the presence of a faint blob near the centre of this map although, it appears to be slightly shifted compared to the $^{12}$CO~\textit{J=3$\rightarrow$2} detection. This central region of emission is detected in $^{12}$CO~\textit{J=1$\rightarrow$0} from the BIMA SONG observations, and the position is closer to the MIPS 24~$\mu$m central detection location. Due to the small area covered by of the BIMA footprint, the northwest bright end region was partly missed, only the southeast bright end was detected.

\begin{figure}
\begin{center}
\subfigure[Northwest bright end region of NGC~2976 at RA~9:47:08.1 Dec~+67:55:50.2 with $\sigma=0.012$K]{
\includegraphics*[width=\columnwidth, trim=15mm 0mm 20mm 10mm, clip]{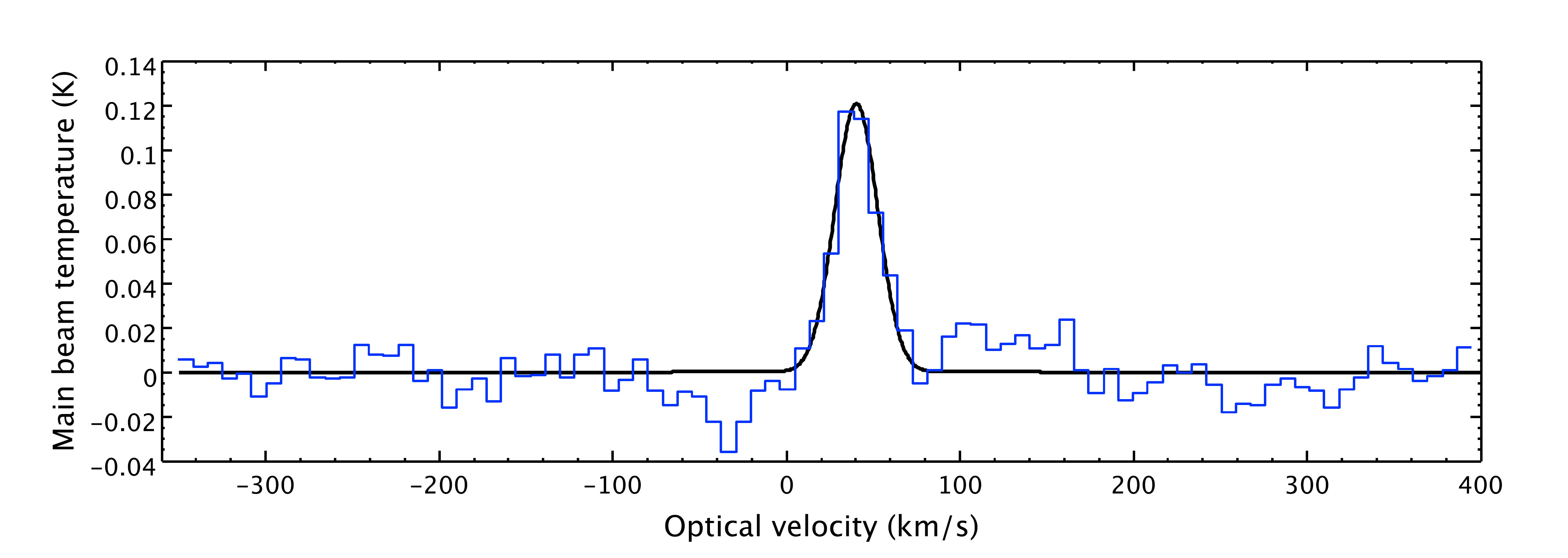}}
\vspace{0.1in}
\subfigure[Southeast bright end of NGC~2976 at RA 9:47:23.6 Dec +67:54:01.2 with $\sigma=0.013$K]{
\includegraphics*[width=\columnwidth, trim=17mm 0mm 20mm 10mm, clip]{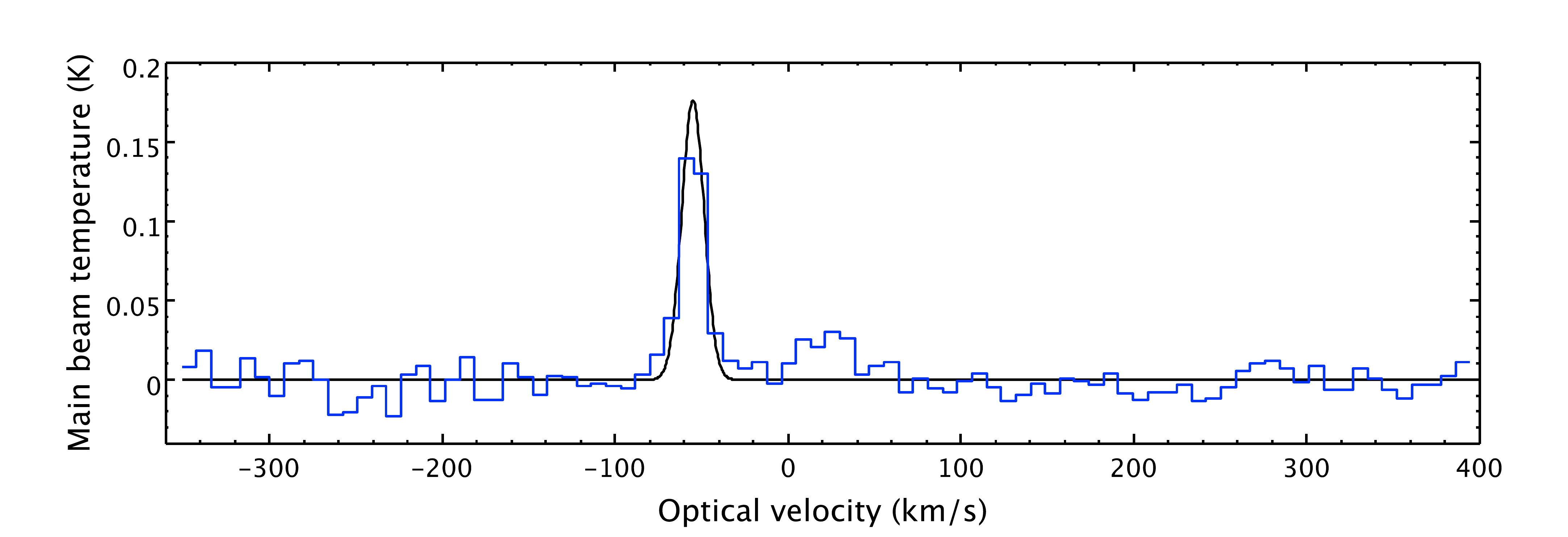}}
\vspace{0.1in}
\subfigure[Central bright region of NGC~3351 at RA 10:43:57.6 Dec +11:42:13.9 with $\sigma=0.012$K]{
\includegraphics*[width=\columnwidth, trim=15mm 0mm 20mm 10mm, clip]{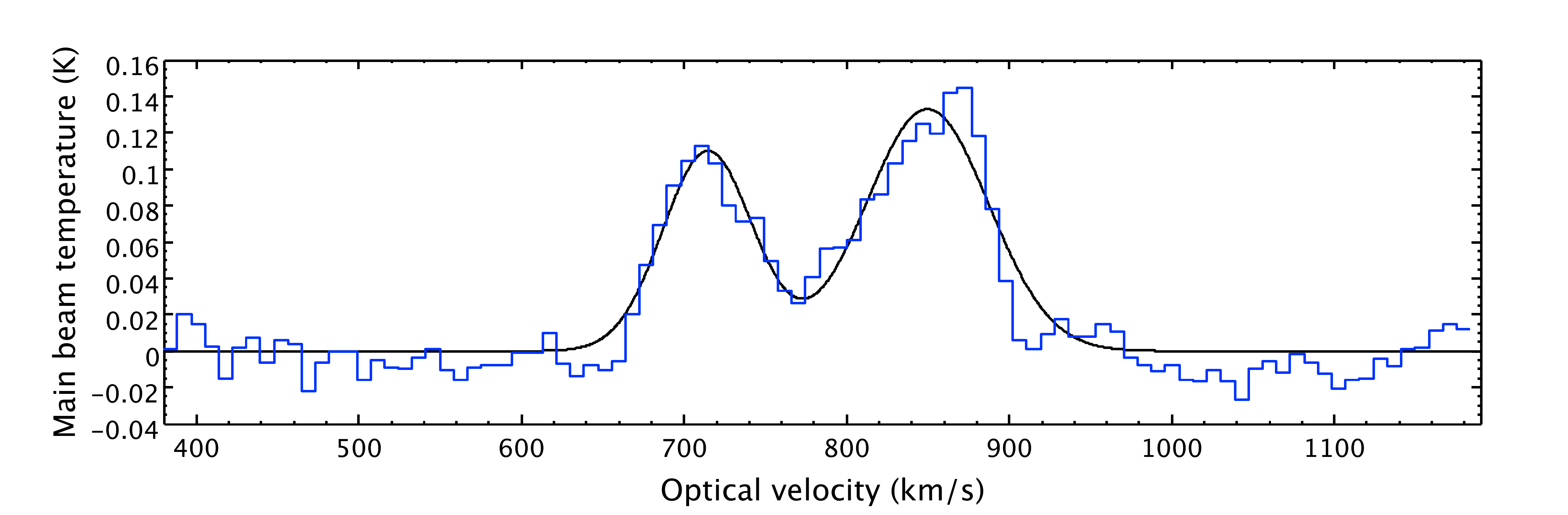}}	
\caption{$^{12}$CO~\textit{J=3$\rightarrow$2} spectra, on the main-beam antenna temperature ($T_{\rm mb}$) scale, of the brightest pixels in NGC~2976 and NGC~3351, smoothed to 10~km~s$^{-1}$. All spectra are overlaid with Gaussian fits. The central velocity of spectrum~(a) is fitted at 40.3~km~s$^{-1}$, with a peak temperature of 0.12~K and FWHM of 30.0~km~s$^{-1}$. Spectrum~(b) has a central velocity of -51.8~km~s$^{-1}$, with a peak temperature of 0.14~K and FWHM of 15.0~km~s$^{-1}$. Spectrum~(c) has peak temperatures of 0.12~K and 0.14~K, with the lower peak velocity at 715~km~s$^{-1}$ and higher peak velocity at 850~km~s$^{-1}$. The FWHM is 72.4~km~s$^{-1}$ for the lower peak and 94.6~km~s$^{-1}$ for the higher peak.}
\label{fig:specline}
\end{center}
\end{figure}

The $^{12}$CO~\textit{J=3$\rightarrow$2} line width of the southeast bright end in NGC~2976 is narrower, with half power line width of around 15~km~s$^{-1}$, compared to the northwest bright end of around 30~km~s$^{-1}$. The spectra of these two bright end regions are shown in Figs.~\ref{fig:specline}~(a) and (b). The peak intensity in $T_{\rm MB}$ is 0.14~K. The emission line can be traced along the inverse-S-like structure from the southeast end towards the northwest end, with the central velocity of the line shifting from --53~km~s$^{-1}$ towards 68~km~s$^{-1}$. The reader is referred to \citet{Wilson:2012} for the velocity field (moment 1) and the velocity dispersion (moment 2) maps of the galaxy.

NGC~3351 has a dominant circumnuclear region detected in all wavebands. The distribution of the $^{12}$CO~\textit{J=3$\rightarrow$2} integrated intensity across the galactic region displays a huge contrast between the dominant centre and the surrounding area. Only a fraction of the area around the southern part of the $2'$ ring is detected in our $^{12}$CO~\textit{J=3$\rightarrow$2} map. The signal from the northern part of the ring is weak and therefore we do not trace the entire ring structure. The bar that is visible in the optical image is not traced in our $^{12}$CO~\textit{J=3$\rightarrow$2} map either.

The complex structure to the southwest of the nucleus, between the centre of the galaxy and the $2'$ ring, only shows up in our $^{12}$CO~\textit{J=3$\rightarrow$2} map. The structure extending from the nucleus towards the southern ring on the east also seems to be offset from the detection region in the NRO $^{12}$CO~\textit{J=1$\rightarrow$0} map. However, this southeast complex and the branch extending slightly towards the northwest, are in fact tracing the dust lane \citep{Swartz:2006} surrounding the nuclear region. These dust lanes are visible in the single filter optical image\footnote{Refer to Fig.~1 in \citet{Swartz:2006}}, and they extend along the leading edge of a bar that is oriented at an angle of $110^{\circ}$ east of north. From the IRAC 3.6~$\mu$m and the optical image, this bar terminates at the $2'$ ring, with faint spiral arms extending beyond the ring. As well as the double ring structure (see Section~\ref{Target Galaxies}), NGC~3351 has an interesting double bar feature, too. An inner bar terminating at the $15''$ ring is detected in the BIMA $^{12}$CO~\textit{J=1$\rightarrow$0} observation \citep{Helfer:2003}. This $15''$ bar is almost perpendicular to the outer $2'$ bar, but is too small to be resolved at the resolution of HARP-B.

The $^{12}$CO~\textit{J=3$\rightarrow$2} emission line near the central region of NGC~3351 exhibits a prominent twin peak feature, as depicted in Fig.~\ref{fig:specline}~(c). This feature points to the rotation of the $15''$ ring and the inflow/outflow of gas within the inner bar. Each peak has a rather broad, $\sim$70~--~90~km~s$^{-1}$ half power line width, with a peak intensity of 0.13~K. 
 
\subsection{$R_{31}$ Line Ratio and Molecular Gas Mass}
\label{CO Line Ratio}

To estimate the H$_{2}$ molecular gas mass using the CO-to-H$_2$ conversion factor ($X_{\rm CO}$), one often uses the $^{12}$CO~\textit{J=1$\rightarrow$0} transition. However, as \citet{Greve:2005} point out, the $^{12}$CO~\textit{J=1$\rightarrow$0} line, which includes emission from the more diffuse molecular gas, does not trace star formation on a one-to-one basis. The $^{12}$CO~\textit{J=3$\rightarrow$2} emission instead correlates almost linearly with the global star formation rate over five orders of magnitude \citep[e.g.,][]{Iono:2009}. Hence, to derive M$_{\rm{H_2}}$ from the warmer and denser gas region, where $^{12}$CO~\textit{J=3$\rightarrow$2} is thermalised, the ratio between $^{12}$CO~\textit{J=3$\rightarrow$2} and $^{12}$CO~\textit{J=1$\rightarrow$0} is important. For this purpose we use the $R_{31}$ map produced using the steps outlined in Section~\ref{Line Ratio}.

\begin{figure}
\centering
\subfigure[Ratio map]{
\includegraphics*[width=3.1in, angle=0]{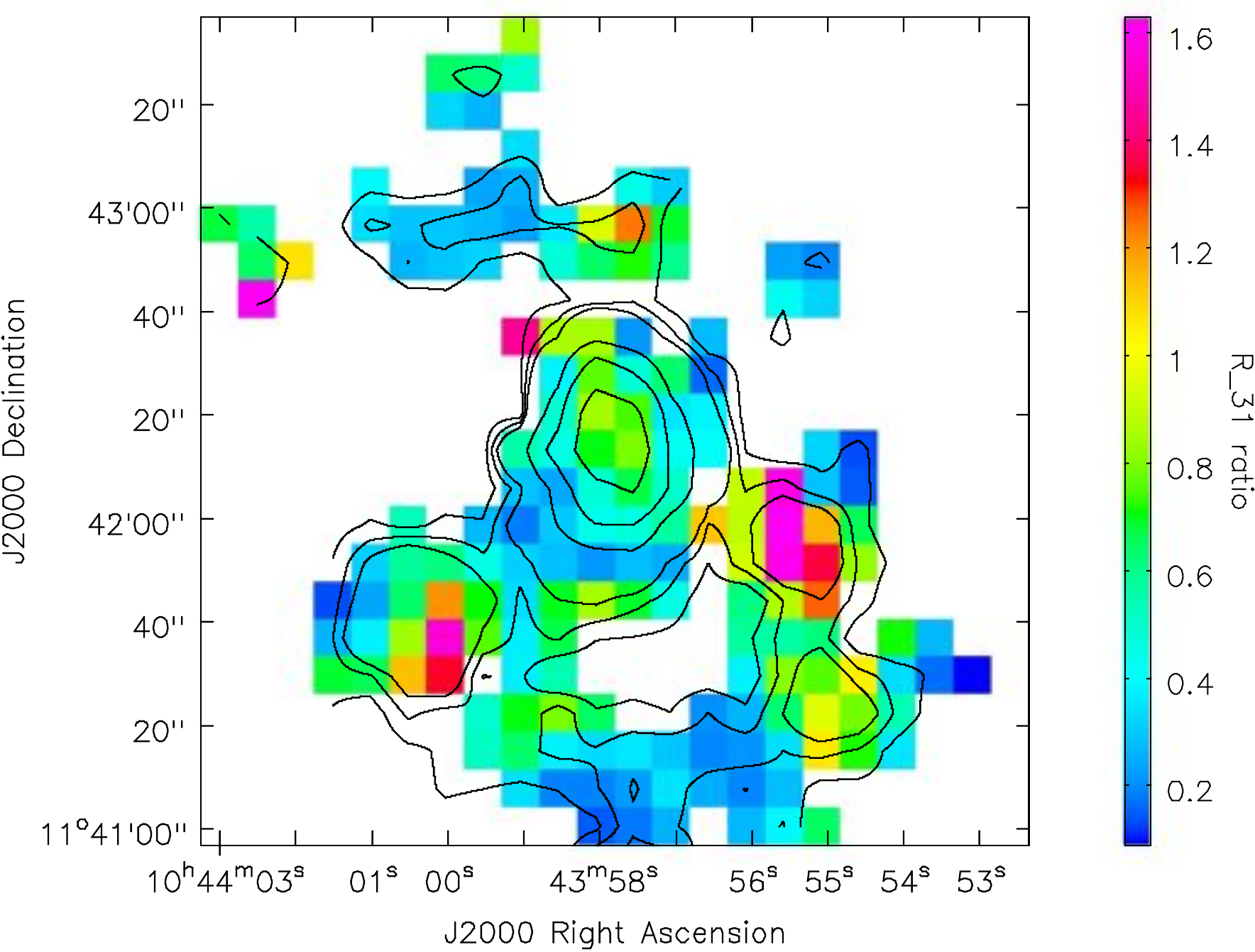}}
\hspace{0.2in}
\subfigure[$R_{31}$ distribution]{
\includegraphics*[width=3.1in, angle=0, trim=0mm 0mm 0mm 0mm, clip]{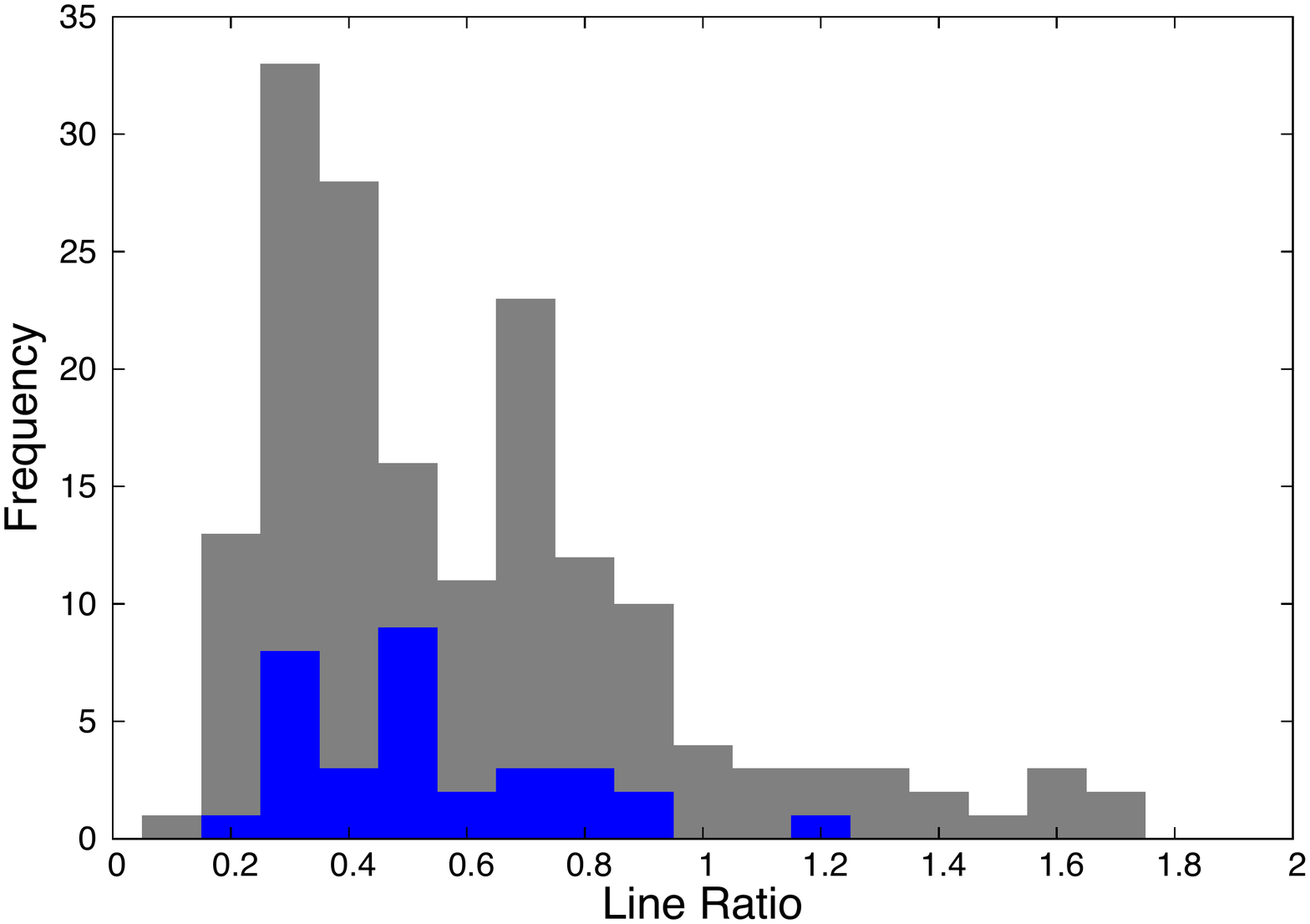}}
\caption{Panel~(a) shows the $R_{31}$ ratio map for NGC~3351 constructed using the $^{12}$CO~\textit{J=1$\rightarrow$0} map from the NRO Atlas. The grey contour line is the $^{12}$CO~\textit{J=3$\rightarrow$2} contour map to show the position of the ratio map relative to the galaxy. Panel~(b) shows the distribution of the value of $R_{31}$ across the galaxy. The grey histogram shows the pixel count when including all the pixels within the ratio map, whereas in the blue histogram we discard pixels within the ratio map with SNR less than 2.}
\label{fig:ratiomap_3351}
\end{figure}

In Fig.~\ref{fig:ratiomap_3351}~(a), we plot the distribution of $R_{31}$ across the disk of NGC~3351. Averaging $R_{31}$ across the entire galaxy disk, we obtain a global mean ratio of 0.49$\pm$0.03. This value agrees within $\sim$20\% with the $R_{31}$ value obtained by comparing our map with the single-beam $^{12}$CO~\textit{J=1$\rightarrow$0} observations, which are obtained from a central pointing on the galaxy. For example, using the $^{12}$CO~\textit{J=1$\rightarrow$0} data from \citet{Sage:1993}, we compute a ratio of 0.48$\pm$0.03 for NGC~3351. The higher $R_{31}$ observed near the complex that extends towards the southwest of the centre of NGC~3351, is the direct consequence of the offset in the detection of $^{12}$CO~\textit{J=3$\rightarrow$2} and $^{12}$CO~\textit{J=1$\rightarrow$0}, as explained in Section~\ref{Results and Discussion}.

\begin{figure*}
\centering
\includegraphics*[width=2\columnwidth, angle=0]{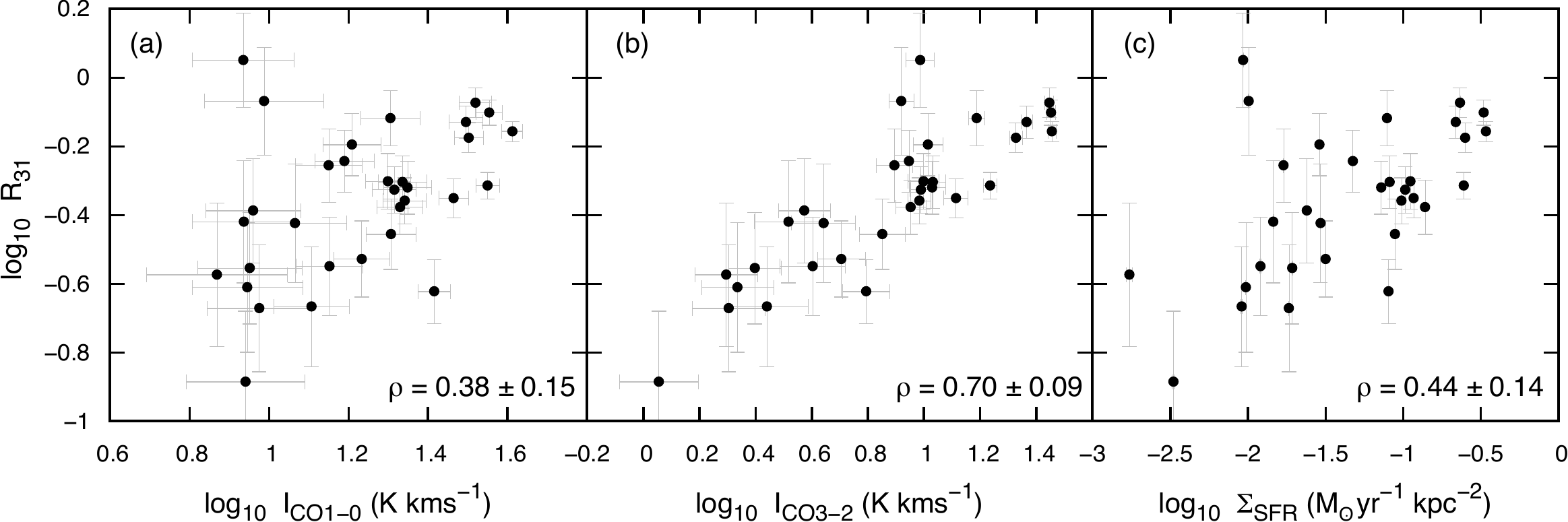}
\caption{Correlation of $R_{31}$ (with SNR$>$2) with $^{12}$CO~\textit{J=1$\rightarrow$0}, $^{12}$CO~\textit{J=3$\rightarrow$2} and SFR surface density respectively, for NGC~3351. The Pearson correlation coefficient ($\rho$) is listed at the bottom left of each panel.}
\label{fig:R31_correlation}
\end{figure*}

Fig.~\ref{fig:ratiomap_3351}~(b) shows the variation of the $R_{31}$ ratio within NGC~3351. The observed range of $R_{31}$ values, mostly between 0.3~--~0.8, is similar to the range of global values obtained for the 28 nearby galaxies studied by \citet{Mauersberger:1999} (0.2~--~0.7), and also similar to the range (0.4~--~0.8) observed in individual giant molecular clouds in M33 \citep{Thornley:1994, Wilson:1997}. We note that similar variations in $R_{31}$ have been seen in other galaxies reported so far from the NGLS survey \citep{Wilson:2009, Warren:2010, Bendo:2010, Irwin:2011, Sanchez:2011}. 

In Fig.~\ref{fig:R31_correlation} we plot $R_{31}$ as a function of $I_{\rm{CO(1-0)}}$, $I_{\rm{CO(3-2)}}$ and the SFR surface density. We see only very weak correlation between the $I_{\rm{CO(1-0)}}$ line brightness and $R_{31}$, indicating a very weak correlation in the spatial distribution of the total molecular gas mass, as traced by $I_{\rm{CO(1-0)}}$. We do, however, note a correlation between $ R_{31}$ and $I_{\rm{CO(3-2)}}$, indicating that the $I_{\rm{CO(3-2)}}$ line is bright where the line ratio is largest. This might also explain the very weak correlation between the $R_{31}$ and the total molecular gas mass traced by $I_{\rm{CO(1-0)}}$. The area with the bright $^{12}$CO~\textit{J=3$\rightarrow$2} line indicates an area with denser gas, hence inevitably increasing the total molecular gas mass traced by $I_{\rm{CO(1-0)}}$, as the less dense molecular gas is gravitationally attracted to the dense region. Plotting $R_{31}$ against the SFR surface density maps in the rightmost panel of Fig.~\ref{fig:R31_correlation} reveals a weak correlation ($\rho=0.44\pm0.14$). It appears that for NGC~3351, the warmer denser gas traced by high $R_{31}$ ratios is reasonably well correlated with the star formation activity, on the spatial scales set by the pixel size of the map ($108.9\times10^3$~\rm{pc}$^2$). 

\begin{figure*}
\centering
\includegraphics*[width=2.0\columnwidth, angle=0]{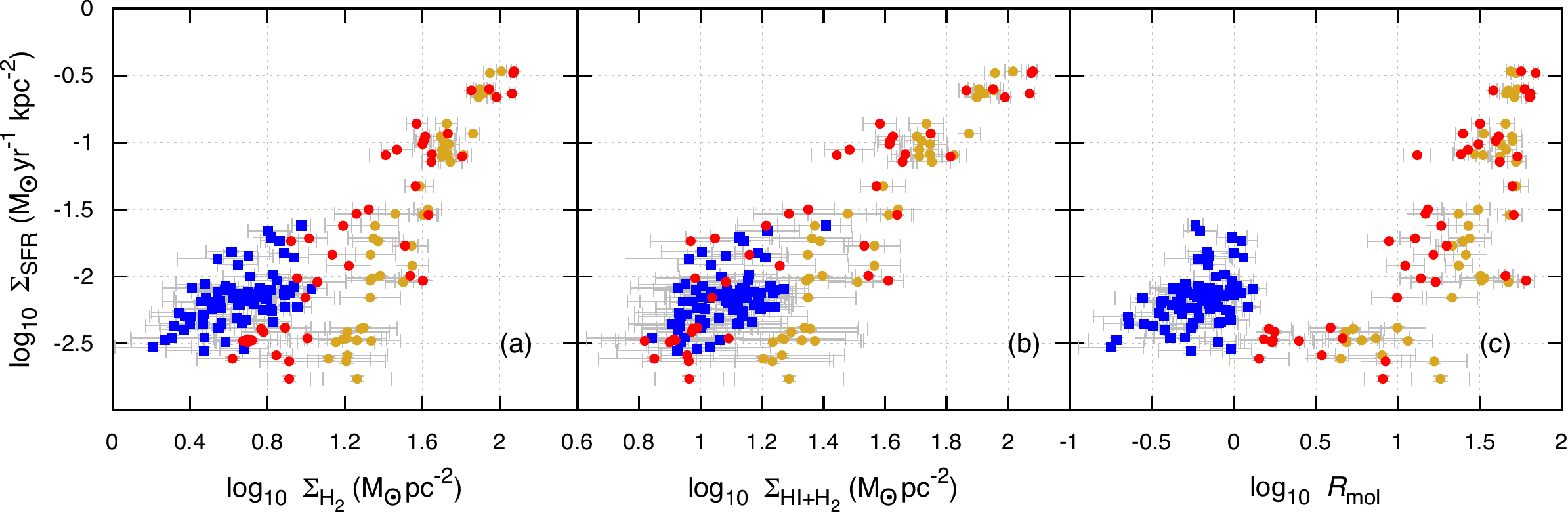}
\caption{Star formation rate density as a function of different gas phases: (left) H$_2$ only, (centre) \textsc{Hi}+H$_2$ and (right) H$_2$-to-\textsc{Hi} ratio. Each point in these diagrams represents a single pixel corresponding to the pixel size of our $^{12}$CO~\textit{J=3$\rightarrow$2} map. Pixels from NGC~2976 are plotted as blue squares, while regions from NGC~3351 are represented by red circles. The H$_2$ gas mass for both of these data sets were derived from our $^{12}$CO~\textit{J=3$\rightarrow$2} data, whereas the yellow diamond points represent the pixels from NGC~3351 in which the H$_2$ gas mass was derived from the NRO $^{12}$CO~\textit{J=1$\rightarrow$0} data.} 
\label{fig:SFR}
\end{figure*}

\begin{figure*}
\centering
\includegraphics*[width=2.0\columnwidth, angle=0]{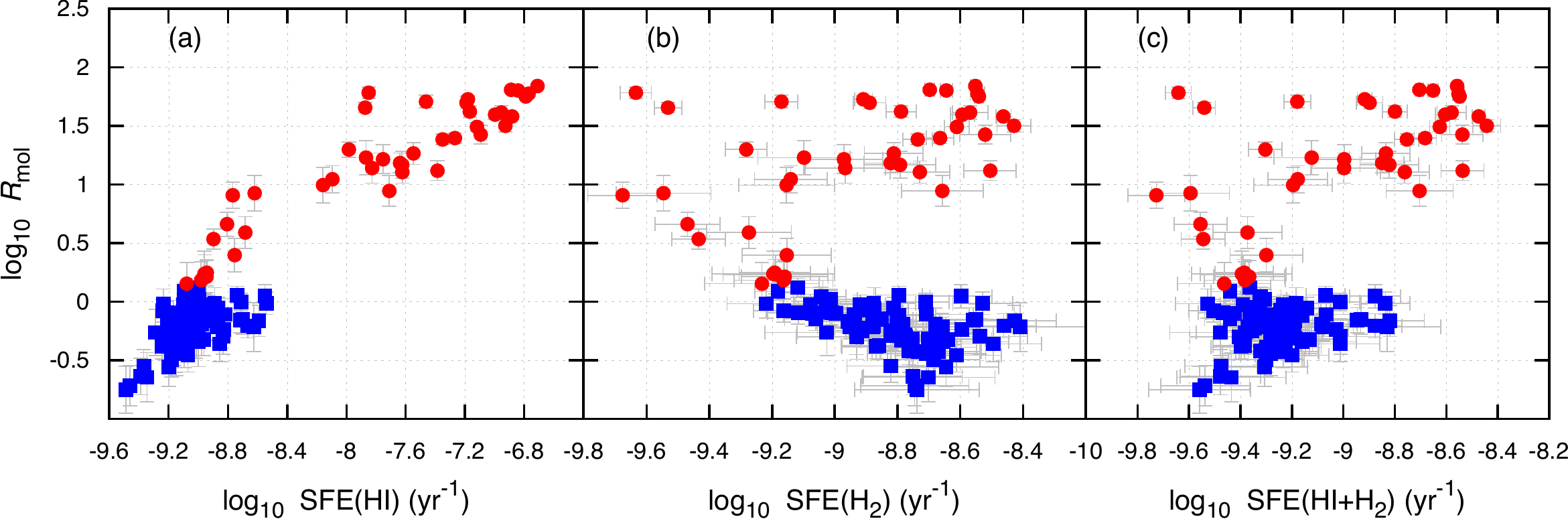}
\caption{$R_{\rm mol}$ as a function of different star formation efficiency: (left) SFE(\textsc{Hi}), (centre) SFE(H$_2$) and (right) SFE(\textsc{Hi}+H$_2$). Each point in these diagrams represents a single pixel corresponding to the pixel size of our $^{12}$CO~\textit{J=3$\rightarrow$2} map. Pixels from NGC~2976 are plotted as blue squares, while regions from NGC~3351 are represented by red circles.} 
\label{fig:Rmol}
\end{figure*}

We next derive M$_{\rm{H_2}}$ using the $R_{31}$ value calculated from the $^{12}$CO~\textit{J=3$\rightarrow$2} and the $^{12}$CO~\textit{J=1$\rightarrow$0} data. The $^{12}$CO luminosity is computed using the following expression:
\begin{equation}
L_{\rm{CO}} = I_{\rm{avg}}\times N_{\rm{pix}}\times23.5\times(D\times\Delta_{\rm{pix}})^2,
\label{eq:lco}
\end{equation}
\noindent where $L_{\rm{CO}}$ is the $^{12}$CO luminosity in K~km~s$^{-1}$~pc$^2$, $I_{\rm{avg}}$ is the average $^{12}$CO intensity (K~km~s$^{-1}$) obtained from the integrated intensity map on main beam temperature scale, $N_{\rm{pix}}$ is the number of pixels included, $D$ is the distance to the galaxy in Mpc and $\Delta_{\rm{pix}}$ is the pixel size in arcseconds. The molecular gas mass is computed using 
\begin{equation}
M_{\rm{H_2}} = 1.6\times10^{-20} \times L_{\rm{CO}} \times X_{\rm{CO}},
\label{eq:mh2_CO1-0}
\end{equation}
\noindent where $L_{\rm{CO}}=L_{CO(3\rightarrow2)}/R_{\rm{31}}$.
In this paper, we use $X_{\rm{CO}}=2 \times 10^{20}$ (K~km~s$^{-1}$)$^{-1}$ \citep{Strong:1988}, consistent with other papers published in the NGLS series, and assume that this conversion factor does not vary across the galactic disk. We do not take into account the effects of metallicity. We note however, that, using the equation reported in \citet{Israel:2000} to calibrate $X_{\rm CO}$ for metallicity, the H$_2$ gas mass estimates for NGC~3351 would remain almost unchanged.

We now investigate the effect of various R$_{\rm{31}}$ values on M$_{\rm{H_2}}$. First, we assume a generic line ratio of 0.6, which is a typical ratio appropriate for the molecular gas in Galactic and extragalactic GMCs \citep{Wilson:2009}. \citet{Israel:2008} also found similar line ratios (0.6$\pm$0.13) from observations of 15 nearby galaxies with modest starbursts. Second, we use the global mean ratio of 0.49 for NGC~3351, calculated directly from the present observations. The M$_{\rm H_{2}}$ derived using the global mean $R_{31}$ value (0.49$\pm$0.03) averaged across the disk of NGC~3351 is (3.3$\pm$0.4) $\times 10^8$ M$_{\odot}$ and agrees well (within the errors) with the value estimated using $R_{31}=0.6$ from \citet{Wilson:2009}, which results in M$_{\rm{H_2}}=$(2.7$\pm$0.3) $\times 10^8$ M$_{\odot}$. Likewise for NGC~2976, using the \citet{Wilson:2009} value, we compute M$_{\rm H_{2}}$ of (0.27$\pm$0.04) $\times 10^8$ M$_{\odot}$ (we have not derived the R$_{\rm{31}}$ value for NGC~2976 due to the lack of calibrated $^{12}$CO~\textit{J=1$\rightarrow$0} data from BIMA as explained previously). These estimates for the warmer gas (based on $^{12}$CO~\textit{J=3$\rightarrow$2} data) are typically lower by a factor of 2~--~3 than the M$_{\rm{H_2}}$ based on lower-$J$ $^{12}$CO~data (e.g., M$_{\rm{H_2}}=$8.14 $\times 10^8$ M$_{\odot}$ for NGC~3351 and M$_{\rm{H_2}}=$0.61 $\times 10^8$ M$_{\odot}$ as reported by \citet{Leroy:2009} using their $^{12}$CO~\textit{J=2$\rightarrow$1} HERACLES data) which traces more diffuse and cooler gas. We thus conclude that our estimates of the warm and denser gas (based on $^{12}$CO~\textit{J=3$\rightarrow$2} data) are fairly insensitive against various $R_{31}$ values.

\subsubsection{Molecular gas mass and $\Sigma_{\rm SFR}$}

In this section we investigate how the ratio H$_{2}/$\textsc{Hi} (which we denote as $R_{\rm mol}$) varies as a function of the $\Sigma_{\rm SFR}$ in two different environments, an \textsc{Hi} rich dwarf (NGC~2976) and an H$_2$ dominated galaxy (NGC~3351). We use the M$_{\rm{H_2}}$ values computed in Section~\ref{CO Line Ratio} based on the generic $R_{31}=0.6$ line ratio. 

In Fig.~\ref{fig:SFR}, we plot the $\Sigma_{\rm SFR}$ as a function of the surface density of H$_2$ ($\Sigma_{\rm H_2}$ in panel~\ref{fig:SFR}~(a)), surface density of total gas mass ($\Sigma_{\textsc{Hi}+{\rm H_2}}$ in panel~\ref{fig:SFR}~(b)), and the ratio of H$_2$-to-\textsc{Hi} ($R_{\rm mol}$ in panel~\ref{fig:SFR}~(c)). In the case where $^{12}$CO~\textit{J=3$\rightarrow$2} has been used to derive M$_{\rm{H_2}}$, the data are shown as red circles and blue squares for NGC~3351 and NGC~2976, respectively. For comparison we also show the distribution of M$_{\rm{H_2}}$ based on $^{12}$CO~\textit{J=1$\rightarrow$0} which is applicable only in the case of NGC~3351 (yellow circles). Each point in these diagrams represents the corresponding quantity calculated within an area of $15.6\times10^3$~\rm{pc}$^2$ for NGC~2976 and $108.9\times10^3$~\rm{pc}$^2$ for NGC~3351 (which is the pixel size of our $^{12}$CO~\textit{J=3$\rightarrow$2} maps). The surface densities of both molecular and atomic gas are calculated directly by dividing the gas mass estimated within that single pixel by the corresponding pixel area, taking into account the inclination of the galaxy. We find no correlation between the \textsc{Hi} surface density ($\Sigma_{\textsc{Hi}}$) and $\Sigma_{\rm SFR}$, hence we have not plotted it here, although we note that most pixels have $\Sigma_{\textsc{Hi}}$ surface densities below 10~$M_{\odot}~\rm{pc}^{-2}$, as seen in \citet{Schruba:2011} and \citet{Leroy:2008}. 

Fig.~\ref{fig:SFR}~(a) shows that a tight correlation exists between $\Sigma_{\rm SFR}$ and H$_2$ surface density based on $^{12}$CO~\textit{J=3$\rightarrow$2} data for NGC~3351 (power law index $n=1.53$). A similar trend is seen in Fig.~\ref{fig:SFR}~(b) where $\Sigma_{\rm SFR}$ is plotted as a function of $\Sigma_{\textsc{Hi}+{\rm H_2}}$ ($n=1.65$). However, when $^{12}$CO~\textit{J=1$\rightarrow$0} data are used (yellow circles), then the slope of the correlation becomes steeper ($n=2.46$ for $\Sigma_{\rm H_2}$ and $n=2.75$ for $\Sigma_{\textsc{Hi}+{\rm H_2}}$, respectively). We suggest that the difference in slopes might be due to the fact that the diffuse areas where $^{12}$CO~\textit{J=1$\rightarrow$0} originates may not be directly related to active star formation. 

In Fig.~\ref{fig:SFR}~(c) we plot $\Sigma_{\rm SFR}$ as a function of $R_{\rm mol}$. This plot clearly shows that NGC~2976 is mainly \textsc{Hi} dominated, whereas NGC~3351 is H$_2$ dominated. The pixel distribution for NGC~3351 shows a trend where higher $R_{\rm mol}$ values correspond to higher values of $\Sigma_{\rm SFR}$. We suggest that this indicates that active star formation takes place in these areas (since there is more available H$_2$ to fuel star formation). No such trend is seen in NGC~2976. The power law relation between $R_{\rm mol}$ and $\Sigma_{\rm SFR}$ is not surprising given the Schmidt-Kennicutt relation \citep{Kennicutt:1989} that links areas of high H$_{2}$ concentration with higher rates of star formation (this relation is also reflected in Fig.~\ref{fig:SFR}~(a)). Nevertheless, one would expect that the higher the H$_2$-to-\textsc{Hi} ratio is, the higher the rate of star formation (i.e., a tight power law relation) if \textsc{Hi} does not contribute to star formation activity. But we do not see a very distinctive evidence here. This might be due to the narrow dynamic range to fully sample the $R_{\rm mol}$ parameter space, or it could potentially show that the role of \textsc{Hi} in star formation cannot be completely ignored \citep{Fumagalli:2008,Leroy:2008,Fumagalli:2009,Glover:2012}. In fact, as shown in Fig.~\ref{fig:Rmol}~(a), higher values of $R_{\rm mol}$ do indeed correlate with higher star formation efficiency SFE(\textsc{Hi}) (SFR surface density per unit \textsc{Hi} gas surface density), especially for NGC~3351. 

However, this linear relationship between SFE(\textsc{Hi}) and $R_{\rm mol}$ is not seen with the SFE(H$_2$) (which is defined as the SFR surface density per unit H$_2$ gas surface density), nor with SFE(\textsc{Hi}+H$_2$) (defined as the SFR surface density per unit neutral gas surface density), as shown in Fig.~\ref{fig:Rmol}~(b) and (c). A further investigation collecting all the SINGS galaxies in NGLS survey would be very instructive in this case.

\subsection{Correlation with PAH Emission}
\label{Correlation with PAH Emission}

\begin{figure*}
\centering
\subfigure[NGC~2976]{
\includegraphics[width=0.95\columnwidth, angle=0]{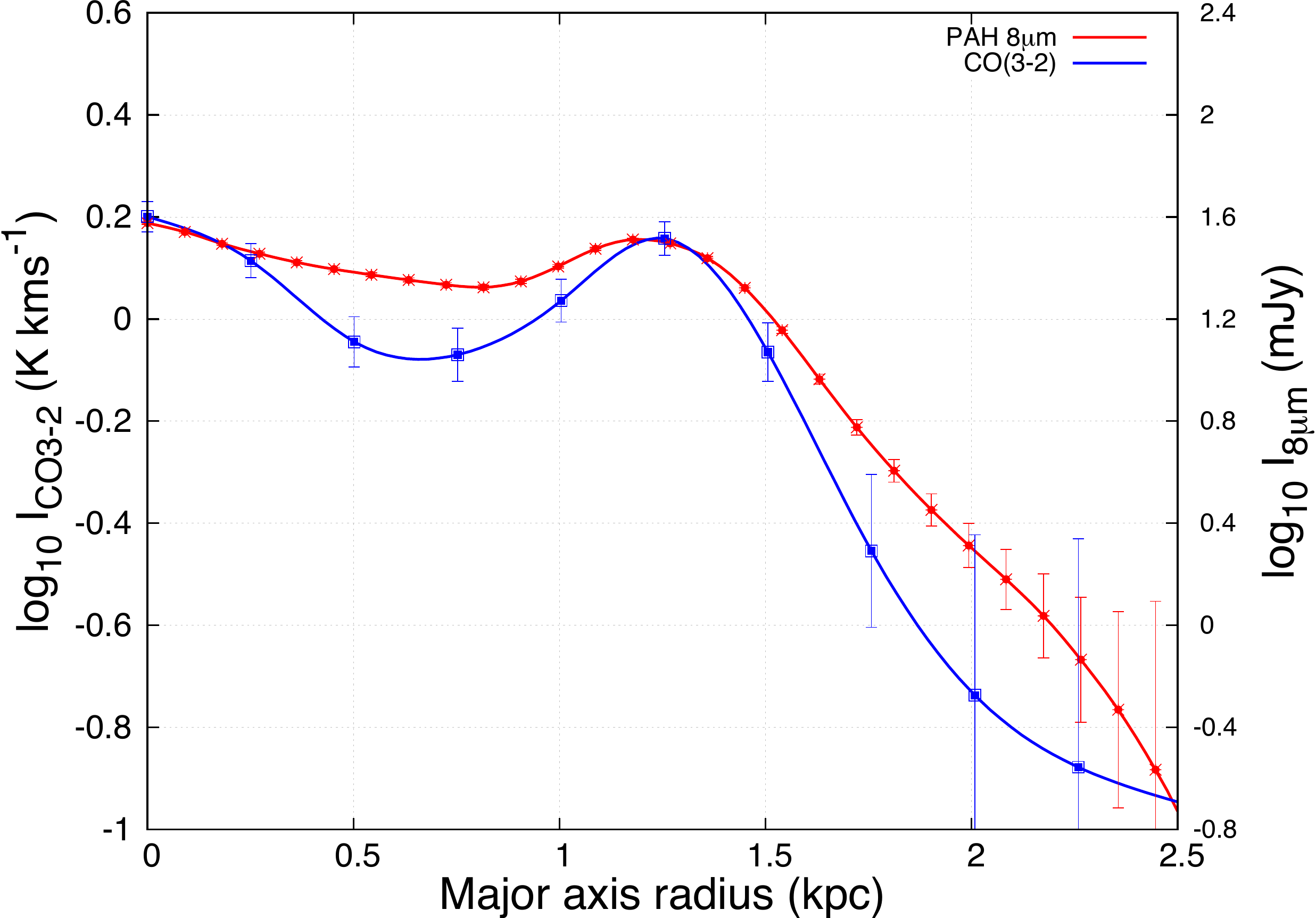}}
\hspace{0.2in}
\subfigure[NGC~3351]{
\includegraphics[width=0.95\columnwidth, angle=0]{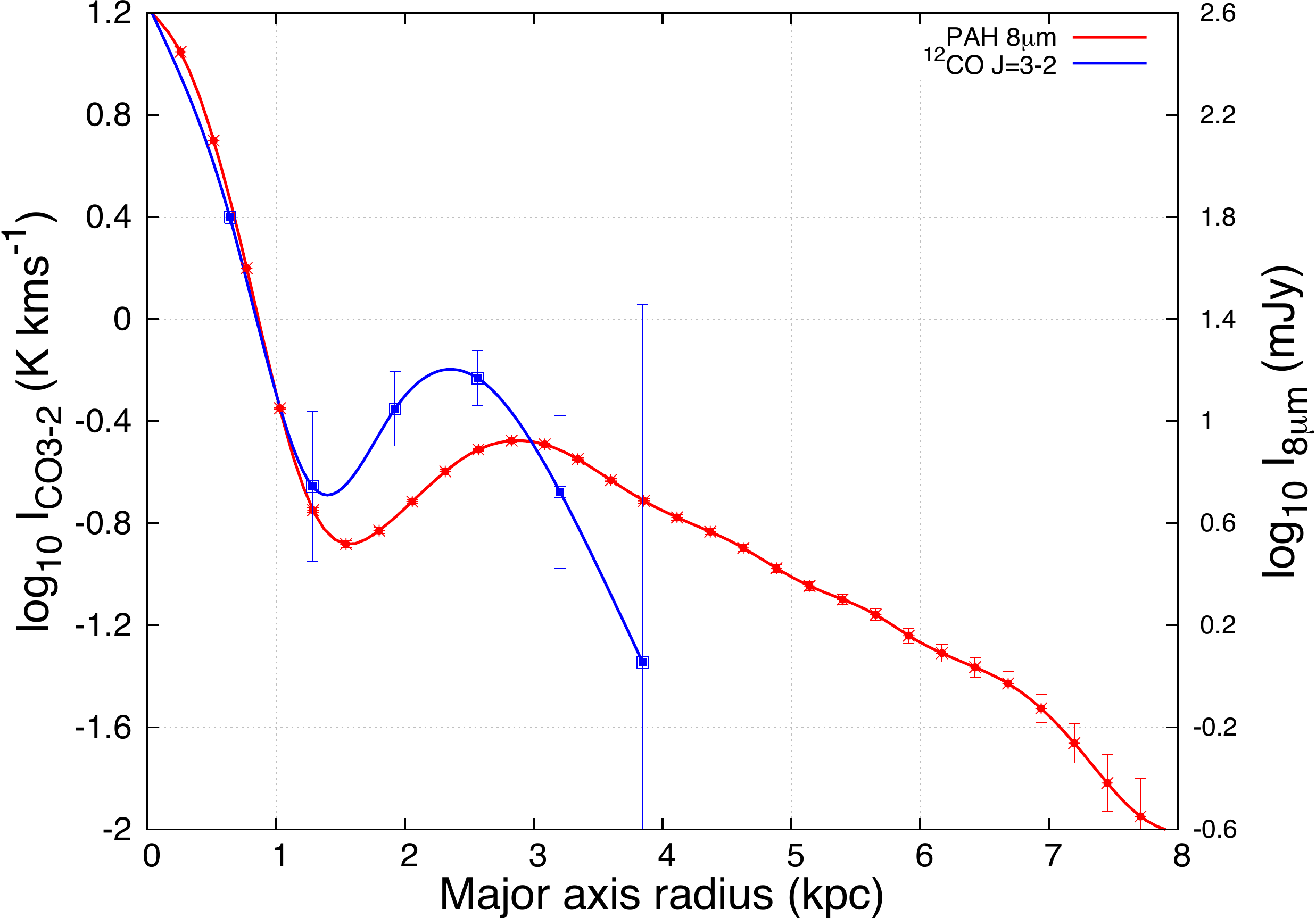}}
\caption{Comparison of the $^{12}$CO~\textit{J=3$\rightarrow$2} radial profile with that of \textit{Spitzer} IRAC 8$\mu$m for both NGC~2976 and NGC~3351. $^{12}$CO~\textit{J=3$\rightarrow$2} intensity is in K~km~s$^{-1}$ (blue, left axis) and PAH 8 $\mu$m is in mJy (red, right axis). All data were convolved with kernels that match the PSF of $^{12}$CO~\textit{J=3$\rightarrow$2} data.}
\label{fig:rprofile}
\end{figure*}

In both Figs.~\ref{fig:4in1_2976} \& \ref{fig:4in1_3351}, the large structures traced by $^{12}$CO~\textit{J=3$\rightarrow$2} appear to match those seen in the PAH 8~$\mu$m image. The two bright end regions of NGC~2976 can be seen in both wavebands. The ratio between the surface brightness of these bright end regions to the centre of the galaxy is larger in the PAH 8~$\mu$m image, and the weak inverse-S shape structure that was visible in the $^{12}$CO~\textit{J=3$\rightarrow$2} map appears to show regions of stronger emission in the PAH 8~$\mu$m map as well. There is, however, a difference in the small scale structures among these images. One example is the small structure directly to the south of the northern bright end region in the $^{12}$CO~\textit{J=3$\rightarrow$2} map (see contour at R.A. 09:47:10 and Dec. +67:55:30 of Fig.~\ref{fig:4in1_2976}) which does not have a comparable counterpart in the PAH 8~$\mu$m image. 

The emission from the central circumnuclear ring region of NGC~3351 dominates the brightness map in all three wavebands. The $2'$ ring structure is clearly seen in the PAH 8~$\mu$m image. This ring structure is only partially detected in $^{12}$CO~\textit{J=3$\rightarrow$2} map. We note that the areas that were detected here do not seem any brighter in the PAH 8~$\mu$m image.  

One might expect that $^{12}$CO~\textit{J=3$\rightarrow$2}, which is excited in the warm and dense molecular gas regions nearer to the star formation sites, would have a high spatial correlation with PAH emission, if the PAH emission is connected to star formation activity. \citet{Regan:2006} studied the radial distribution of the 8~$\mu$m emission and the $^{12}$CO~\textit{J=1$\rightarrow$0} emission for 11 disk galaxies and found a high spatial correlation between them. \citet{Bendo:2010} compared the radial profiles of PAH 8~$\mu$m surface brightness to the $^{12}$CO~\textit{J=3$\rightarrow$2} in NGC~2403 and found that the scale lengths in both cases are statistically identical. But their examination in sub-kpc scale regions within the galaxy revealed that $^{12}$CO~\textit{J=3$\rightarrow$2} and PAH 8~$\mu$m surface brightness seem to be uncorrelated. Here we further investigate this correlation for NGC~2976 and NGC~3351 by comparing the radial profile of the $^{12}$CO~\textit{J=3$\rightarrow$2} emission to the PAH 8~$\mu$m surface brightness.   

Fig.~\ref{fig:rprofile} shows the radial profile of PAH 8~$\mu$m and the $^{12}$CO~\textit{J=3$\rightarrow$2} surface brightness for both NGC~2976 and NGC~3351. The general shapes appear to match in both galaxies; however on smaller spatial scales the agreement is less apparent.

The scale lengths, defined as the radial distances where the intensity drops by $1/e$ from the peak intensity, for NGC~2976 are 1.65~kpc and 1.85~kpc for the $^{12}$CO~\textit{J=3$\rightarrow$2} and PAH 8~$\mu$m emission, respectively. These scale lengths are larger than the two bright end regions which correspond to the local maxima at $\sim$1.3~kpc away from the centre of the galaxy. Beyond $\sim$1.3~kpc, both emissions decay rapidly with almost the same rate. The ratio between the peak brightness of the two bright end regions with the galaxy centre, and the position of these bright end regions, are similar in both radial profiles. This might suggest that within high surface brightness regions, both components correlate better compared to lower brightness regions. \citet{Bendo:2010} examined this relation in NGC~2403 at the sub-kpc scale, and found a similar trend. In their plot\footnote{Fig.~14 in \citet{Bendo:2010}.}, it appears that both indicators only overlapped in the area where $^{12}$CO~\textit{J=3$\rightarrow$2} intensity is higher than 1.0~K~km~s$^{-1}$ and PAH 8~$\mu$m surface brightness is higher than 2.0~MJy~sr$^{-1}$.

Within the central $\sim$1~kpc of NGC~3351, the two radial profiles follow each other closely; this is likely to be due to the resolution limit imposed by the HARP-B beam size (FWHM of $\sim$660~pc at a distance of 9.33~Mpc). The PSF for both data have been convolved to the same resolution and the galaxy centre contains a bright source that is unresolved at the resolution of 14.5$''$. Away from the central region, the two profiles do not trace each other well anymore. Note that the ratio of the $2'$ ring-to-nucleus intensity of the $^{12}$CO~\textit{J=3$\rightarrow$2} is higher than the PAH 8~$\mu$m profile. Outside the $2'$ ring, the $^{12}$CO~\textit{J=3$\rightarrow$2} emission decays much faster than the PAH 8~$\mu$m emission,the latter extending almost twice as far as the $^{12}$CO~\textit{J=3$\rightarrow$2} profile. As we saw in NGC~2976, this might indicate that the correlation of the two components is better in higher brightness region in the galaxy. But again, the detection of $^{12}$CO~\textit{J=3$\rightarrow$2} in this region is too faint to reach a definite conclusion.

\begin{figure}
\centering
\includegraphics*[width=\columnwidth, angle=0]{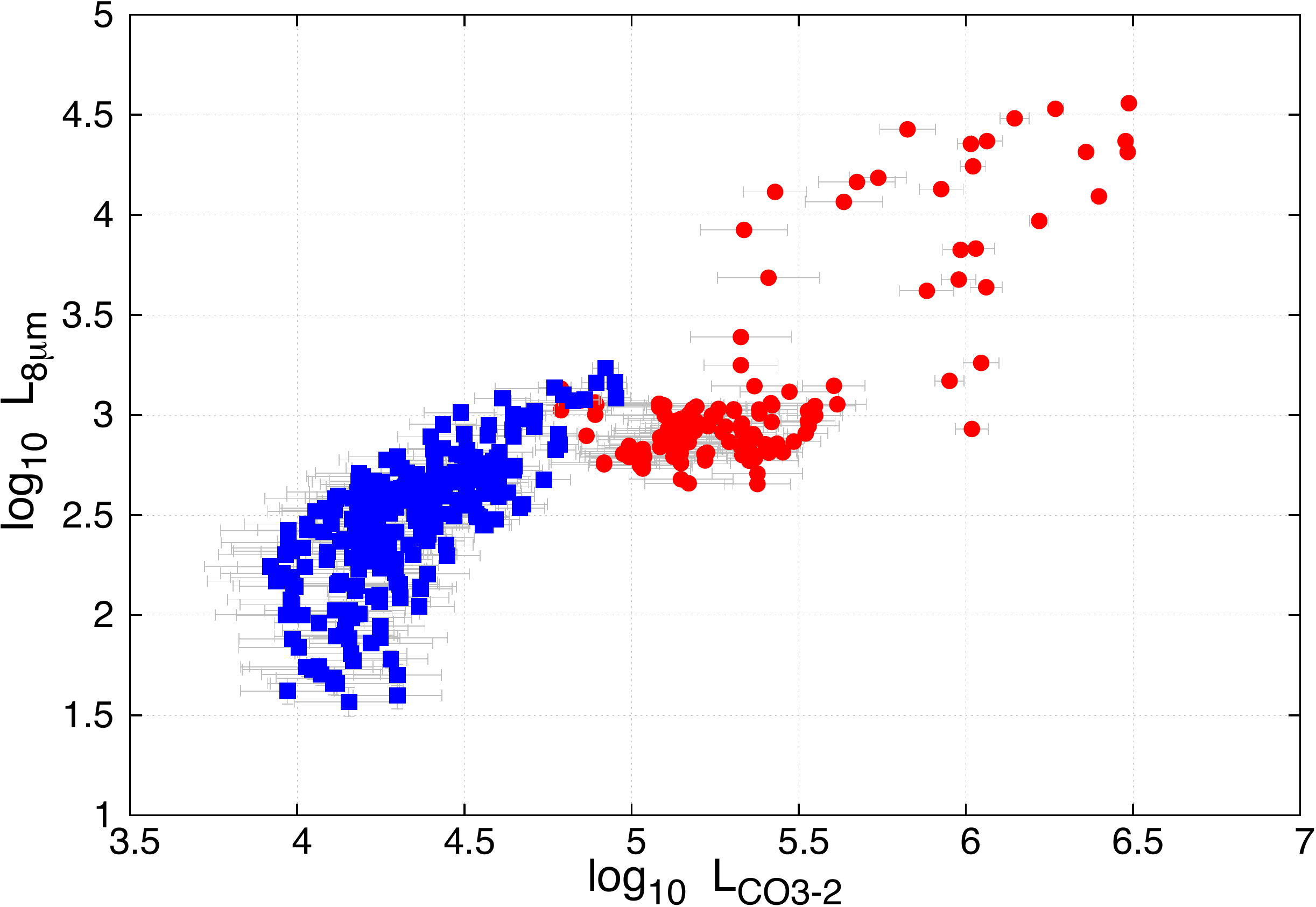}
\caption{Pixel-by-pixel comparison of the $^{12}$CO~\textit{J=3$\rightarrow$2} luminosity with the PAH 8~$\mu$m luminosity. Each red circle represent a pixel from NGC~3351, while each blue square corresponds to a single pixel in NGC~2976 map.}
\label{fig:CO3-2_8um}
\end{figure}

To further examine this relation, we compare both detections pixel by pixel and plot their correlation in Fig.~\ref{fig:CO3-2_8um}, using only the pixels detected within our $^{12}$CO~\textit{J=3$\rightarrow$2} map. Again, the PAH 8~$\mu$m map is convolved and re-gridded to match the resolution of our $^{12}$CO~\textit{J=3$\rightarrow$2} map. Note that due to the difference in distance, each point within this plot corresponds to an area of $15.6\times10^3$~pc$^2$ for NGC~2976 and $108.9\times10^3$~pc$^2$ for NGC~3351. As with the radial profile comparison, this pixel-by-pixel approach seems to show that both kinds of emission are quite well correlated globally. However, there seems to be a larger scatter at the higher surface brightness end of the plot, and the correlation seems to break down at $\log L_{\rm CO3\rightarrow2} < 4.5$. This seems to agree with the argument from \citet{Bendo:2010} that this correlation would break down at lower intensity or smaller spatial scales. It would be interesting to further examine this relation with a larger sample of galaxies to determine where this correlation begins to fail.

\begin{figure}
\centering
\includegraphics*[width=\columnwidth, angle=0]{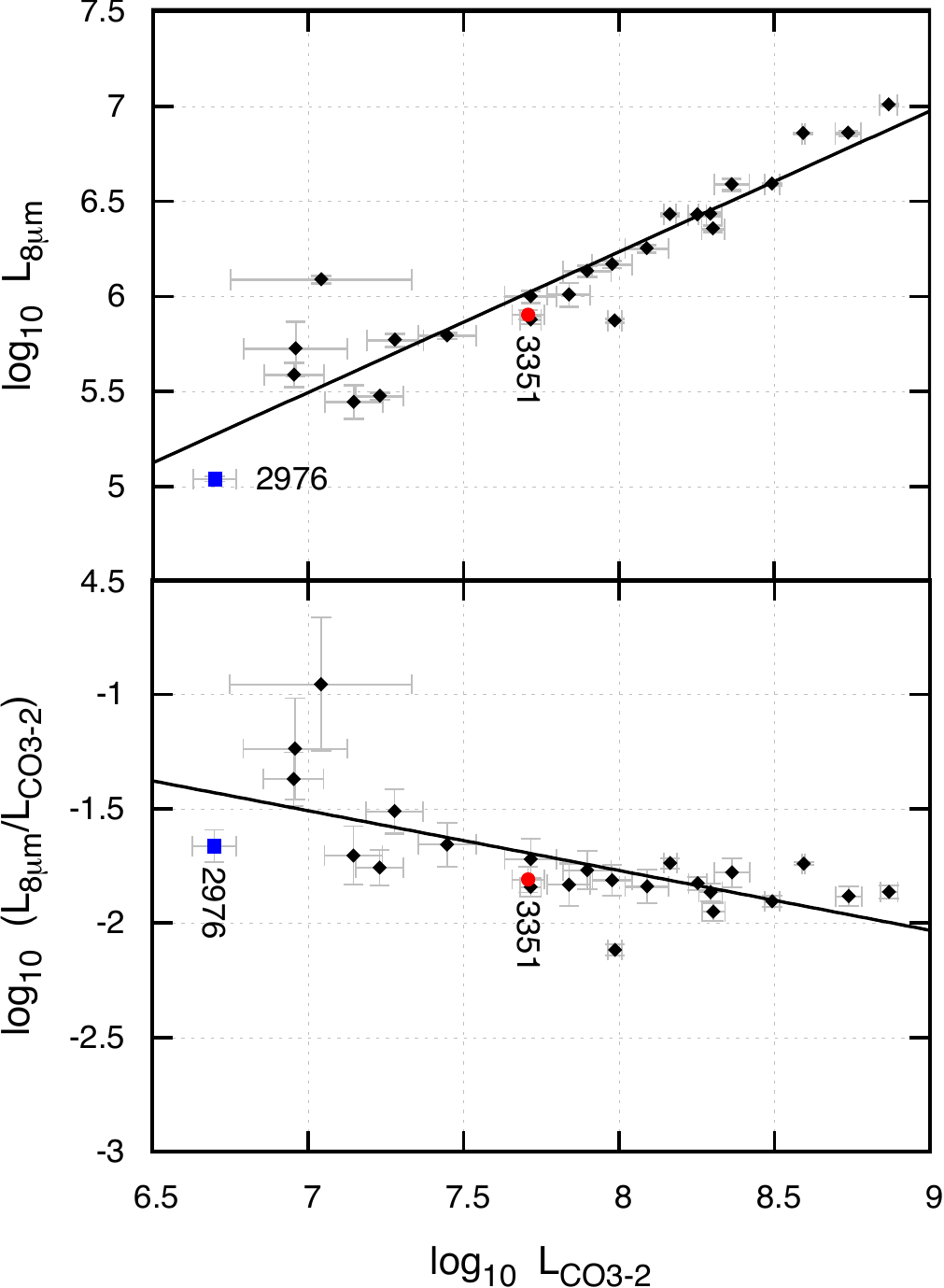}
\caption{Correlation of total integrated PAH 8$\mu$m luminosity and the total integrated $^{12}$CO~\textit{J=3$\rightarrow$2} luminosity of the 25 SINGS galaxies in the NGLS samples.}
\label{fig:PAH_CO_correlation}
\end{figure}

In order to determine whether this relationship only holds for these two galaxies, we examined all the SINGS galaxies within the NGLS to see if NGC~2976 and NGC~3351 are typical of the entire sample. The other field and Virgo galaxies within the NGLS samples were not included here, as they lack the corresponding PAH 8~$\mu$m data from SINGS survey. In Fig.~\ref{fig:PAH_CO_correlation}, we plot the \textit{total} $^{12}$CO~\textit{J=3$\rightarrow$2} luminosity against the \textit{total} PAH 8~$\mu$m luminosity for NGC~2976 and NGC~3351, together with 23 SINGS galaxies that have $^{12}$CO~\textit{J=3$\rightarrow$2} detected. The luminosity values are tabulated in Table~\ref{tab:PAHvsCO}. The $^{12}$CO~\textit{J=3$\rightarrow$2} maps of these 23 galaxies are obtained from the early release of the NGLS data set \citep{Wilson:2012}. The PAH 8 $\mu$m maps are obtained from the same source as described in Section~\ref{Archival Data}. 

Clearly, the correlation of the two indicators is tight. There seems to be an approximately linear relation between the two, with $\log L_{\rm 8~\mu m} \approx (0.74\pm0.06)\log I_{\rm CO(3-2)} - 0.32$. The Pearson correlation coefficient is as high as 0.97, with both NGC~2976 and NGC~3351 lying close to the best fit line. To investigate whether this tight relation is not entirely dominated by the strong nuclear emission regions, we made a similar plot of the median luminosity instead of total luminosity, and found that the high correlation still holds.

Even though both $^{12}$CO~\textit{J=3$\rightarrow$2} and PAH 8~$\mu$m are believed to be linked to star formation, their origins within the ISM and their fundamental excitation mechanisms are different \citep[e.g.,][]{Young:1991, Tielens:2008}. The 8~$\mu$m emission originates from PAHs found near the stars, in particular in the photospheres of AGB stars \citep{Tielens:2008}, while molecular CO is formed on the surface of interstellar dust grains. The surface brightness of PAH 8~$\mu$m emission is mostly proportional to the number of FUV (far ultraviolet) photons and the number density of PAH molecules. The $^{12}$CO~\textit{J=3$\rightarrow$2} intensity, on the other hand, is controlled mainly by the collision rate with the molecular hydrogen, the temperature of the gas and the density of molecular hydrogen. It is thus important to understand why they correlate so well in a global environment.

\begin{table}
\caption{\textit{Total} luminosity of PAH 8 $\mu$m and $^{12}$CO~\textit{J=3$\rightarrow$2} of 25 SINGS galaxies within NGLS samples.}
\begin{center}
\begin{tabular}{l c c c}
\hline
Galaxy		& Distance	& $L_{\rm PAH 8\mu m}$		& $L_{\rm CO(3-2)}$			\\[0.5ex]
			& (Mpc)		& (MJy pc$^2$)			& ($\times 10^7 $K~km~s$^{-1}$ pc$^2$)		\\[0.5ex]
\hline
NGC~0628		& 7.3		& 0.99 $\pm$ 0.07		& 5.2 $\pm$ 1.0		\\
NGC~0925		& 9.1		& 0.38 $\pm$ 0.15		& 0.9 $\pm$ 0.2		\\
NGC~2403		& 3.2		& 0.30 $\pm$ 0.04		& 1.7 $\pm$ 0.3		\\
NGC~2841		& 14.1		& 1.22 $\pm$ 0.04		& $<$ 1.1		\\
NGC~2976		& 3.6		& 0.11 $\pm$ 0.03		& 0.50 $\pm$ 0.08		\\
NGC~3031		& 3.6		& 0.53 $\pm$ 0.33		& 0.91 $\pm$ 0.35		\\
NGC~3034		& 3.6		& 7.12 $\pm$ 0.01		& 39.2 $\pm$ 0.4		\\
NGC~3049		& 22.7		& 0.26 $\pm$ 0.21		& 1.4 $\pm$ 0.3		\\
NGC~3184		& 11.1		& 1.46 $\pm$ 0.04		& 9.5 $\pm$ 1.4		\\
NGC~3198		& 13.7		& 1.01 $\pm$ 0.14		& 6.9 $\pm$ 1.1		\\
NGC~3351		& 9.3		& 0.79 $\pm$ 0.06		& 5.1 $\pm$ 0.6		\\
NGC~3521		& 7.9		& 2.53 $\pm$ 0.01		& 17.9 $\pm$ 1.2		\\
NGC~3627		& 9.4		& 3.86 $\pm$ 0.01		& 31.1 $\pm$ 1.7		\\
NGC~3938		& 14.7		& 1.65 $\pm$ 0.05		& 12.3 $\pm$ 2.0		\\
NGC~4254		& 16.7		& 10.06	 $\pm$ 0.02		& 73.8 $\pm$ 4.8		\\
NGC~4321		& 16.7		& 7.14	 $\pm$ 0.02		& 54.7 $\pm$ 5.3		\\
NGC~4559		& 9.3		& 0.58 $\pm$ 0.08		& 1.9 $\pm$ 0.4		\\
NGC~4569		& 16.7		& 2.25 $\pm$ 0.04		& 20.1 $\pm$ 1.7		\\
NGC~4579		& 16.7		& 1.34 $\pm$ 0.07		& 7.9 $\pm$ 1.4		\\
NGC~4631		& 7.7		& 2.67 $\pm$ 0.02		& 14.6 $\pm$ 0.7		\\
NGC~4736		& 5.2		& 0.75 $\pm$ 0.05		& 5.2 $\pm$ 0.4		\\
NGC~4826		& 7.5		& 0.74 $\pm$ 0.02		& 9.7 $\pm$ 0.5		\\
NGC~5033		& 16.2		& 3.60 $\pm$ 0.07		& 23.1 $\pm$ 3.0		\\
NGC~5055		& 7.9		& 2.75 $\pm$ 0.02		& 19.7 $\pm$ 1.7		\\
UGC~05720	& 25.0		& 0.58 $\pm$ 0.04		& 2.8 $\pm$ 0.6		\\
\hline
\end{tabular}
\label{tab:PAHvsCO}
\end{center}
\end{table}

One possible explanation is that there is a spatial co-existence of the PAH molecules and molecular CO in photo-dissociation regions (PDR), although not at specific sub-kpc locations within individual clouds. At the front of a PDR, the UV flux from hot stars or the interstellar radiation field excites the PAH molecules and heats the small grains in the ISM. This is responsible for the generation of intense radiation from PAHs and the continuum dust emission in the mid-infrared. CO rotational transitions are generally believed to be generated from regions that reside deeper in the photo-dissociation regions \citep{Hollenbach:1999}. In this region, the UV source is still strongly photo-dissociating species like OH and H$_2$O, but is weaker compared to the ionisation front, and thus molecular hydrogen and CO start to form in this region. Given a certain amount of UV photons, a higher density of PAHs would absorb a larger portion of these stellar fluxes, and reduce the photo-dissociation of CO molecules, hence resulting in a positive correlation of the two constituents. Also, both PAHs and molecular CO would not exist in high UV radiation region, as both species would be dissociated.

The same scenario can be applied to the star forming regions within the molecular clouds. If the $^{12}$CO~\textit{J=3$\rightarrow$2} intensity indicates the amount of gas fuelling star formation in the core, it could be approximately proportional to the number of UV photons radiated by these newly born O and B stars. If a linear fraction of these photons is absorbed by the PAHs, their surface brightness will then be approximately proportional to the $^{12}$CO~\textit{J=3$\rightarrow$2} intensity.

There are other alternative explanations in the literature, for example the association between the PAH emission and the cold dust \citep{Haas:2002, Bendo:2008}. If molecular gas is associated with cold dust, then the PAH and $^{12}$CO~\textit{J=3$\rightarrow$2} would be expected to be associated with each other as well. In general, we find that the $^{12}$CO~\textit{J=3$\rightarrow$2} luminosity does correlate well with PAH 8~$\mu$m luminosity at large scales, but this correlation might not hold for smaller sub-kpc scale.

\section{Summary \& Conclusions}
\label{Summary and Conclusions}

We have presented $^{12}$CO~\textit{J=3$\rightarrow$2} maps of NGC~2976 and NGC~3351, obtained using HARP-B on the JCMT. We compared our observations to the optical, $^{12}$CO~\textit{J=1$\rightarrow$0}, PAH 8~$\mu$m, \textsc{Hi} and $\Sigma_{\rm SFR}$ maps constructed using a combination of 24~$\mu$m and FUV data. $^{12}$CO~\textit{J=3$\rightarrow$2} emission from NGC~2976 was strong at both ends of the galaxy's major axis, whereas in NGC~3351 the emission peaks in the nuclear region. NGC~2976 showed a large scale structure that was seen only in the $^{12}$CO~\textit{J=2$\rightarrow$1} image, but not in other wavebands included in this paper, due to the coarse resolution of the $^{12}$CO~\textit{J=3$\rightarrow$2} and $^{12}$CO~\textit{J=1$\rightarrow$0} maps. In contrast, the dominant circumnuclear region in NGC~3351 was visible in all waveband maps presented here. However, the prominent $2'$ ring structure was only weakly detected in our $^{12}$CO~\textit{J=3$\rightarrow$2} map.

We combined our $^{12}$CO~\textit{J=3$\rightarrow$2} data with $^{12}$CO~\textit{J=1$\rightarrow$0} data from various sources to derive the $R_{31}$ line ratio (for NGC~3351). The ratio values we obtained were within the ranges derived from various nearby galaxy surveys. We then computed M$_{\rm{H_2}}$ using the derived $R_{31}$ as well as the generic value of 0.6 from \citet{Wilson:2009}. We found that M$_{\rm{H_2}}$ estimates are robust against the value of R$_{31}$ used. 

We further examined the correlation between $\Sigma_{\rm SFR}$ and surface density of H$_2$ mass and neutral hydrogen gas mass. We found that the $\Sigma_{\rm SFR}$ correlates better with the $^{12}$CO~\textit{J=3$\rightarrow$2} emission. We used the \textsc{Hi} data from THINGS to derive the H$_2$-to-\textsc{Hi} ratio and compared it with the SFE(\textsc{Hi}). We found that although NGC~2976 is \textsc{Hi} dominated and NGC~3351 is an H$_2$-dominated galaxy, they both show a correlation between $R_{\rm mol}$ and SFE(\textsc{Hi}), which implies that the role of \textsc{Hi} in star formation cannot be excluded. 

We have also studied the correlation of the $^{12}$CO~\textit{J=3$\rightarrow$2} and PAH 8~$\mu$m surface brightness as both are prominent indicators linked to star formation activity. We first investigated their relation within NGC~2976 and NGC~3351 using both the radial distribution of the surface brightness and a pixel-per-pixel comparison. We found that they correlate well in the higher surface brightness regions. We further studied the global correlation of the \textit{total} luminosity of the two physical parameters using the data from all 25 SINGS galaxies within the NGLS and reached a similar conclusion. We suggest that this could be the result of the coexistence of both $^{12}$CO and PAH molecules in the PDR regions, provided that the lifetimes of both species are matched. In conclusion, we suggest that this correlation is high at large spatial scales, but at smaller sub-kpc scales, the correlation may break down.

\section*{Acknowledgements}
The D.Phil. study of B. K. Tan at the University of Oxford was supported by the Royal Family of Malaysia through the King Scholarship Award. The work of J. Leech was supported by a Post Doctoral Research Assistantship from the UK Science \& Technology Facilities Council (STFC). The research work of G. J. Bendo was funded by the STFC, and C. D. Wilson and B. E. Warren were supported by grants from NSERC (Canada). J. H. Knapen acknowledges financial support to the DAGAL network from the People Programme (Marie Curie Actions) of the European Union's Seventh Framework Programme FP7/2007-2013/ under REA grant agreement number PITN-GA-2011-289313. This research has made use of the NASA/IPAC Extragalactic Database (NED) which is operated by the Jet Propulsion Laboratory, California Institute of Technology, under contract with the National Aeronautics and Space Administration. The travelling grants to the JCMT for observations are supported by STFC. We would like to thank Bruce Draine and Pauline Barmby for helpful discussions. The authors would also like to thank all of the staff at the JCMT for their support in carrying out our observations. We would also like to thank the reviewer for all the constructive suggestions.

\subsection*{James Clerk Maxwell Telescope}
The James Clerk Maxwell Telescope is a 15 meter single-dish submillimetre telescope operated by the Joint Astronomy Centre (JAC) on behalf of the Science and Technology Facilities Council (STFC) of the United Kingdom, the Netherlands Organisation for Scientific Research, and the National Research Council of Canada. It is located at Mauna Kea, Hawai'i.


\def\reff@jnl#1{{\rm#1\/}}
\def\aj{\reff@jnl{Astron.~J.}}          
\def\araa{\reff@jnl{Ann.~Rev.~Astron.~Astrophys}}   
\def\apj{\reff@jnl{Astrophys.~J.}}      
\def\apjl{\reff@jnl{Astrophys.~J.~Lett.}}  
\def\apjs{\reff@jnl{Astrophys.~J.~Suppl.~Ser.}} 
\def\ao{\reff@jnl{Appl.Opt.}}           
\def\aap{\reff@jnl{Astron.~Astrophys.}} 
\def\aapr{\reff@jnl{Astron.~Astrophys.~Rev.}} 
\def\aaps{\reff@jnl{Astron.~Astrophys.~Suppl.}} 
\def\baas{\reff@jnl{Bull.~Am.~Ast.~Soc.}} 
\def\expast{\reff@jnl{Exp.~Astron.}}    
\def\mnras{\reff@jnl{Mon.~Not.~R.~Ast.~Soc.}} 
\def\pasp{\reff@jnl{Pub.~Astron.~Soc.~Pac.}} 
\def\pra{\reff@jnl{Phys.Rev.A}}         
\def\prb{\reff@jnl{Phys.Rev.B}}         
\def\prc{\reff@jnl{Phys.Rev.C}}         
\def\prd{\reff@jnl{Phys.Rev.D}}         
\def\prl{\reff@jnl{Phys.Rev.Lett}}      
\newcommand\nat{\reff@jnl{Nature}}      
\def\procspie{\reff@jnl{Proc.~SPIE}}    
\def\josa{\reff@jnl{J.~Opt.~Soc.~Am.}}   

\bibliographystyle{mn2e}
\bibliography{JCMT_NGLS_IX}

\bsp

\label{lastpage}

\end{document}